\documentclass[number,twocolumn,5p]{elsarticle}
\usepackage{amssymb}
\usepackage{amsmath}
\usepackage{siunitx}
\DeclareSIUnit\neq{n_{eq}}

\usepackage[dvipsnames]{xcolor}
\usepackage[unicode]{hyperref}
\usepackage{multirow}
\usepackage{pdflscape} 
\usepackage{tabularray}
\UseTblrLibrary{booktabs,
                siunitx,
                }

\begin{document}

\begin{frontmatter}

    \title{Review of prototypes developed in a \SI{65}{\nano\meter} CMOS imaging technology in view of vertexing applications at a future lepton collider}

    \tnotetext[t2]{This work has been performed in the framework of the OCTOPUS project within the DRD3 Collaboration.}
    
    %Please add your name here, if you have contributed to the paper
    \author[desy]{Finn King\corref{corresAuthor}}
		\cortext[corresAuthor]{Corresponding author}
		\ead{finn.king@desy.de}
    \author[ETH]{Matthew Lewis Franks}
    \author[desy]{Yajun He}
    \author[desy]{Gianpiero Vignola}
    \author[desy]{Simon Spannagel}
    \author[ETH]{Malte Backhaus}
    \author[IPHC]{Auguste Besson}
    \author[cern]{Dominik Dannheim}
    \author[IPHC]{Andrei Dorokhov}
    \author[desy]{Ingrid-Maria Gregor}
    \author[IPHC]{Fadoua Guezzi-Messaoud}
    \author[desy]{Lennart Huth}
    \author[UZH]{Armin Ilg}
    \author[FNSPECTU]{Zdenko Janoska}
    \author[FNSPECTU]{Monika Kuncova}
    \author[UZH]{Anna Macchiolo}
    \author[IPHC]{Frédéric Morel}
    \author[desy]{Sara Ruiz Daza}
    \author[MBI]{Roberto Russo}
    \author[desy]{Judith Schlaadt}
    \author[IPHC]{Serhiy Senyukov}
    \author[FNSPECTU]{Peter	Švihra}    
    \author[desy]{Anastasiia Velyka}
    \author[NikHef]{Håkan Wennlöf}

    \affiliation[cern]{
        organization={CERN},
        addressline={Esplanade des Particules 1},
        city={CH-1211 Geneva},
        country={Switzerland}
    }
   
    \affiliation[desy]{
        organization={Deutsches Elektronen-Synchrotron DESY},
        addressline={Notkestr. 85},
        city={22607 Hamburg},
        country={Germany}
    }
    \affiliation[ETH]{
        organization={ETH Zurich},
        addressline={8092},
        city={Zurich},
        country={Switzerland}
    }
    \affiliation[IPHC]{
        organization={Université de Strasbourg, CNRS, IPHC UMR},
        addressline={7178},
        city={Strasbourg},
        country={France}
    }
    \affiliation[UZH]{
        organization={Universität Zürich},
        addressline={Rämistrasse 71},
        city={8006 Zurich},
        country={Switzerland}
    }
    \affiliation[FNSPECTU]{
        organization={Department of Physics, FNSPE CTU in Prague},
        addressline={Brehova 78/7},
        city={Prague, 115 19},
        country={Czechia}
    }
    \affiliation[MBI]{
        organization={Marietta-Blau-Institut für Teilchenphysik},
        addressline={Dominikanerbastei 16},
        city={1010 Wien},
        country={Austria}
    }
    \affiliation[NikHef]{
        organization={Nikhef},
        addressline={Science Park 105},
        city={Amsterdam, 1098XG},
        country={The Netherlands}
    }
    
    \begin{abstract}
The OCTOPUS project addresses the development and characterization of monolithic active pixel sensors in the TPSCo \SI{65}{\nano\meter} ISC technology in view of vertexing applications at a future lepton collider. Meeting the corresponding requirements---outlined in the ECFA detector road map---will necessitate the simulation, design, and testing of prototypes and a demonstrator chip in this very process.

This work reviews the literature on existing prototypes, summarizing their design characteristics, properties, and performance in charged-particle detection, and provides an overview of previous simulation efforts. The presented results suggest the feasibility of the endeavor while showcasing challenges, the need for further investigations, and providing a foundation for imminent design choices.
    \end{abstract}
    

\end{frontmatter}

%\linenumbers

\journal{Nuclear Instrumentation and Methods in Physics Research A}

%%%%%%%%%%%%%%%%%%%%%%%%%%%%%%%%%%%%%%%%%%%%%%%%%%%%%%%%%%%%%%%%%%%%%%%%%%%%%%%%%%%%%%%%%%%%%%%%%
\tableofcontents

\newpage

\section{Introduction}
\label{intro}

Commercial CMOS imaging sensor technologies allow integrating the sensor and readout in the same silicon substrate, forming Monolithic Active Pixel Sensors (MAPS)~\cite{first_maps1, first_maps2}. Compared to the hybrid detector technology~\cite{hybrids}, MAPS offer several advantages, including a lower material budget because the sensor and readout electronics are integrated into the same chip, as well as a reduced production cost and effort since no sensor-to-chip bonding process is required~\cite{kolanoski_wermes}. MAPS are separated into two types: those produced in a high-voltage CMOS process~\cite{hvcmos} and those with a small collection electrode. The use of a small collection electrode results in a lower pixel capacitance, which offers the possibility to reduce power consumption \cite{snoeys_cap}. In the past decade, MAPS have been successfully used in High-Energy Physics (HEP) detectors~\cite{ultimate2016, alpide2016}. They are expected to play a more important role in future HEP experiments, especially for vertex and tracking systems, whose detectors come with unparalleled combinations of requirements~\cite{ecfa2021}.

In recent years, the HEP community has started qualifying the TPSCo \SI{65}{\nano\meter} ISC technology\footnote{This refers to the \SI{65}{\nano\meter} CIS (CMOS imaging sensor) technology offered by TPSCo (Tower Partners Semiconductor Company Limited).} in the context of the ALICE ITS3 project~\cite{tdr_its3}. Compared to the formerly used larger feature size processes, this process offers the potential of lower power consumption, e.g., due to smaller supply voltages of the digital circuitry~\cite{wsnoeys_summary}, and higher densities of circuit elements.

The OCTOPUS (Optimized CMOS Technology for Precision in Ultra-thin Silicon) project~\cite{octopus_web}, which covers the simulation, development, and evaluation of MAPS implemented in the TPSCo \SI{65}{\nano\meter} ISC technology, targets the vertex-detector requirements of future lepton colliders~\cite{octopus}. This includes a spatial hit resolution below \SI{3}{\micro\meter}, a temporal hit resolution of about \SI{5}{\nano\second}, a material budget equivalent to about \SI{50}{\micro\meter} of silicon, and a power consumption below \SI{50}{\milli\watt\per\square\centi\meter}. In addition, the sensor will have to tolerate the corresponding exposure to ionizing radiation, for example, a neutron equivalent fluence of $\mathcal{O}(\SI{e14}{\neq\per\square\centi\meter})$ and a total ionizing dose of $\mathcal{O}(\SI{100}{\kilo\gray})$ assuming 4 years Z pole operation at FCC-ee~\cite{fccee}. This paper summarizes the results from a large variety of prototypes in the TPSCo \SI{65}{\nano\meter} ISC technology in regard to these requirements and will allow making well-understood design choices and identifying challenges to overcome.

\section{Sensor layouts}
\label{process}

\color{black}
\begin{figure*}[ht]
    \centering
    \includegraphics[width=1\linewidth]{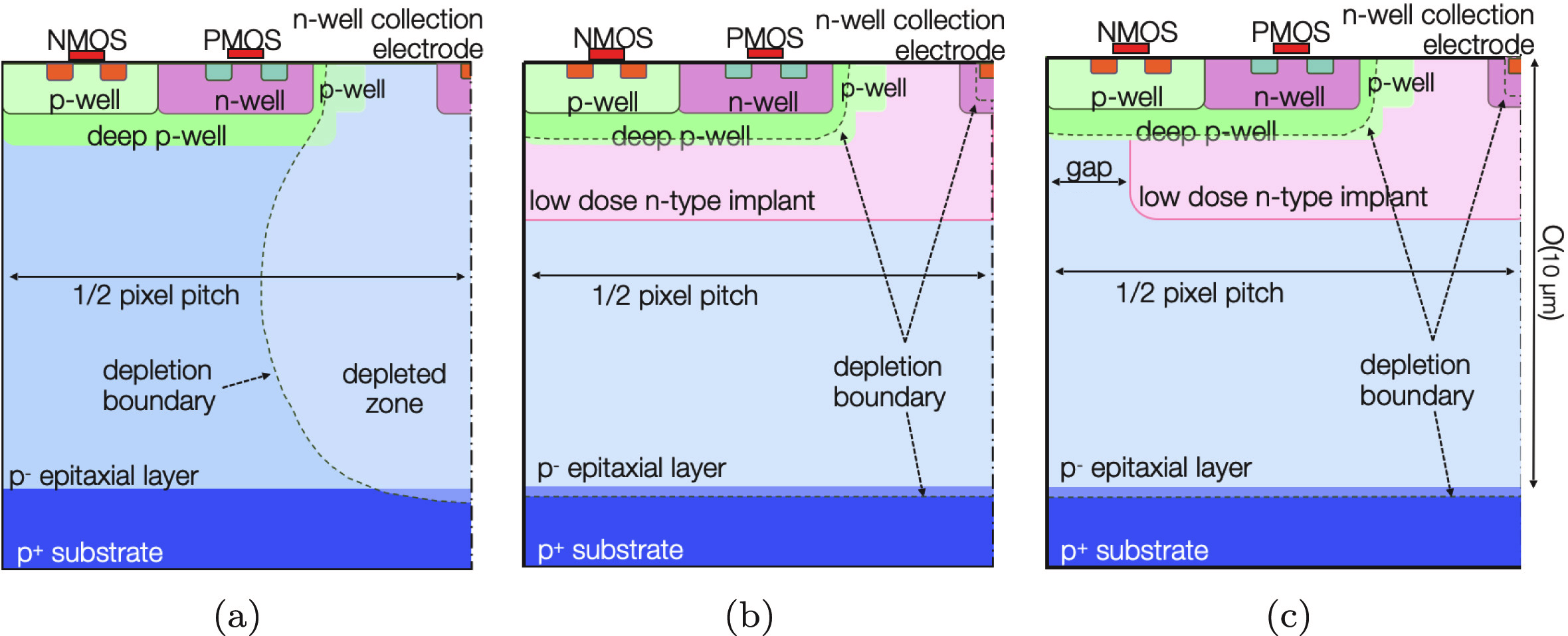}
    \caption{Cross-section of different pixel sensor layouts: (a) the standard layout, (b) the n-blanket layout and (c) the n-gap layout. Modified from \cite{apts}.}
    \label{fig:layout}
\end{figure*}

The considered sensors are manufactured on wafers that feature a high resistivity, p-doped epitaxial layer, with a thickness of $\mathcal{O}(\SI{10}{\micro\meter})$, which is grown on a low resistivity p-doped substrate. The integrated circuitry for signal processing features NMOS and PMOS transistors in their corresponding wells, in turn placed in a deep p-well to isolate them from the sensitive volume. The charge is collected on an n-well electrode. As shown in Figure~\ref{fig:layout}, there are three different sensor layouts, mainly characterized by different charge collection properties, as introduced in~\cite{Snoeys:2023hF}:
\paragraph{Standard layout} The standard layout is the simplest to manufacture and shown in Figure~\ref{fig:landau}(a). The depletion region develops from the collection electrode and reaches the p-substrate layer, taking a balloon-like shape. The volume around the pixel boundary and parts of the volume below the deep p-well remain undepleted, which means that charge transport in these regions is governed by diffusion.
\paragraph{N-blanket layout} This layout is achieved through a process modification which adds a deep, low-dose n-type implant, shown in Figure~\ref{fig:layout}(b). This blanket implant extends the depletion across the full pixel area, as a deeper planar p-n junction forms. However, the lateral electric field in the depletion region is small, except in the vicinity of the collection electrode. This results in slow charge collection from the pixel boundary, particularly if the pixel size is large.
\paragraph{N-gap layout} This modification introduces a gap in the low-dose n-type implant at the pixel boundaries, as shown in Figure~\ref{fig:layout}(c). This introduces a lateral doping gradient and hence an electric field, such that charge carriers created near the pixel boundaries are quickly swept towards the pixel center, suppressing diffusion into neighboring pixels and reducing drift-time variations between different positions inside a pixel.
\\ \\ Note that there are other layouts that lead to an improved lateral field, including, for example, an additional p-type implant below the deep p-well~\cite{Munker_2019}. Simulations show that the additional p-implant leads to a similar charge collection time as the n-gap layout for a Minimum-Ionizing Particles (MIPs) incident at the pixel corner. Both increase the total amount of collected charge by at least a factor of three compared to the n-blanket layout after irradiation~\cite{Munker_2019}. Sensors with this additional p-implant have been investigated in the \SI{180}{\nano\meter} technology within the mini-MALTA~\cite{mini-malta} project and demonstrate a performance similar to the n-gap layout.

In addition to these layouts, there are several more subtle variations of design parameters to fine-tune the properties of these sensors, which have been realized and studied on the Analog Pixel Test Structures (APTSs)~\cite{apts} prototypes:
\paragraph{Doping concentrations} The doping profiles of the various implants are confidential. Variations of the doping concentrations of these implants have been produced on different wafers and compared in~\cite{apts,Snoeys:2023hF}, identifying the most promising configuration among them (referred to as `split 4'). It adjusts the doping concentration of the deep p-well to minimize the impact of the in-pixel circuitry on the electric field in the sensitive volume. The pixel capacitance of the chosen configuration is shown to be smaller or equal to other configurations, depending on the bias conditions. The prototypes discussed in this work employ this `split 4' configuration.
\paragraph{Geometrical parameters} Also, the geometry and size of both the n-well collection electrode and the p-well (e.g. the addition of so-called p-well "fingers") have been studied on prototypes with an n-gap layout and a pixel pitch (distance between the collection electrodes) of \SI{20}{\micro\meter}~\cite{apts}. The different geometric variants do not show a large impact on the sensor properties, except for the size of the n-well collection electrode. A larger n-well collection electrode causes a larger pixel capacitance, with a negative impact on the sensor performance.
\\ \\ There are further optimizations of the sensor design in consideration. For example, other than the three existing sensor layouts, adjusting the size of the gap in the n-gap layout, rectangular or hexagonal pixel geometries, or a staggered arrangement of rows of pixels\footnote{This matrix geometry resembles a brick wall, such that every cell has 6 instead of 8 direct neighbors to mimic a hexagonal pixel geometry}. Some of these have been studied in simulation, such as the gap size and hexagonal pixels, as shown in~\cite{maps_sim_wennolf}. Some have already been realized in prototypes, but systematic studies are not public at the time of writing.

\section{Selected prototypes}
\label{prototypes}
This paper summarizes the results from a set of prototypes produced in the TPSCo \SI{65}{\nano\meter} ISC technology~\cite{Snoeys:2023hF}. These prototypes were produced in two submissions in the context of the CERN EP R\&D project~\cite{EPRnD} and the ALICE ITS3 upgrade~\cite{tdr_its3}.

% MLR1
The first submission in 2020, is referred to as MLR1, and included the Analog Pixel Test Structure (APTS)~\cite{apts, apts-timing}, Digital Pixel Test Structure (DPTS)~\cite{dpts}, Circuit Exploratoire \SI{65}{\nano\meter} (CE65)~\cite{ce65_vci, apts_and_ce65}, Charge Sensitive Amplifier (CSA) test chip (DESY-MLR1)~\cite{desy-mlr1,desy-mlr12}, and additional blocks such as bandgap reference circuits, digital-to-analog converters, temperature diodes as well as structures to test transistors and measure single-event upset cross-sections. A more detailed description of these circuits can be found in~\cite{tdr_its3}. The submission served as testbed for building blocks of a full ASIC towards application of the process in HEP. 

% ER1
A second submission in 2023, called ER1, contained a second version of CE65 (CE65v2)~\cite{ce65v2_iworid,ce65v2_pixel}, the Hybrid-to-Monolithic prototype (H2M)~\cite{h2m}, a second iteration of the CSA test chip (DESY-ER1)~\cite{desy_er1}, the MOnolithic Stitched Sensor (MOSS)~\cite{moss}, and the MOnolithic Stitched Sensor with Timing (MOST)~\cite{most}. The main characteristics and key properties of these prototypes are discussed below, and summarized in Table~\ref{tab:overview}. 

\paragraph*{APTS} 
The APTS employs an analog readout scheme where each pixel in the $4\times4$ matrix is buffered to dedicated analog output pads either by a two-stage source-follower (SF) buffer (implemented partially in-pixel, partially in the periphery)~\cite{apts}, or by an operational amplifier (referred to as the APTS-OA variant~\cite{apts-timing}). This limits the complexity of the chip operation and allows, to some extent, to infer information about the temporal development of the signal. These properties render the chip an excellent candidate to study the impact of the many design options realized in different versions of the chip. This includes various combinations of pixel pitches of 10, 15, 20 and \SI{25}{\micro\meter} and the three sensor layouts discussed above.

\paragraph*{DPTS} 
The DPTS is a more complex chip, featuring a larger matrix of $32 \times 32$ pixels, utilizing only the n-gap sensor layout. It performs several functions in-pixel: resetting the sensor diode, amplification of the generated charge, and digitization of the amplified signal via a discriminator. The discriminator output is binary, but the width of the pulse is proportional to the number of collected electrons in a Time-over-Threshold (ToT) measurement. In addition, pixels experiencing excessively high hit rates (indicating noise) can be selectively masked from readout. The characterization of this chip allows gaining experience in the operation of the contained CMOS circuitry. In addition, both, APTS and DPTS, were studied in regard to their tolerance with respect to ionizing and non-ionizing radiation, and timing characteristics.

\paragraph*{CE65 and CE65v2}
CE65 names a family of chips, designed with the goal to explore charge collection in sensors produced in this technology. The CE65 family contains four chip versions, combining a pitch of 15 or \SI{25}{\micro\meter} with either the standard or n-gap layout. The CE65 matrix contains three regions employing different amplification schemes: a SF stage, a DC-coupled amplifier, and an AC-coupled amplifier, to evaluate different amplifier designs~\cite{ce65_vci}. The second iteration of the chip, CE65v2, consists of $48~\times~24$ pixels adopting this capacitively (AC) coupled preamplifier for the entire matrix. This allows to apply a relatively high reverse bias voltage, above \SI{10}{\volt}, to the sensor. The chip also covers an even larger number of pixel pitches and sensor layouts than CE65: standard, n-blanket, and n-gap; 15, 18, and \SI{22.5}{\micro\meter}; and square and hexagonal matrix geometry~\cite{ce65v2_iworid}.

\paragraph*{DESY-MLR1 and DESY-ER1} 
The two CSA test chips DESY-MLR1~\cite{desy-mlr1,desy-mlr12} and DESY-ER1~\cite{desy_er1} feature fast charge sensitive amplifiers with a Krummenacher-type feedback~\cite{krum_1991}, used in test circuits and a small matrix of $2 \times 2$ pixels. The second iteration of the chip employs enhancements to the CSA, and features direct access to analog signal waveforms generated by the CSA and the output of the discriminator~\cite{desy_er1} to enable detailed study of the amplifier response, and a larger pitch of $35 \times \SI{25}{\square\micro\meter}$, which allows further insights into effects observed in H2M.

\paragraph*{H2M}
The H2M (Hybrid-to-Monolithic) test chip architecture is ported from a hybrid readout chip~\cite{h2m} and features a $64\times16$ pixel matrix. It uses the same fast, analog front end as the DESY-ER1 chip~\cite{desy_er1}, a CSA with Krummenacher feedback. The versatile digital in-pixel circuitry allows the output signal to be compared to a global threshold controlled by an 8-bit DAC. Pixels also include a 4-bit trimming DAC to tune the threshold for each pixel to account for pixel-to-pixel variations. Like DPTS, it features also logic for masking pixels. Both the analog and digital components mandate the relatively large pitch of \SI{35}{\micro\meter}. The chip is operated in a frame-based mode, and the in-pixel circuitry can be operated in four readout modes: ToT, time-of-arrival (ToA, 10 ns binning), hit counting, and a triggered mode. It serves to study the challenges arising with this porting process, and as a test bench for a compact digital cell library \cite{h2m_meas}.

\paragraph*{MOSS and MOST}
The MOSS~\cite{moss,moss_proceeding} and MOST~\cite{most} prototypes are tailored towards the ALICE upgrade ITS3, and their characterization is still in progress at the time of writing. They represent the largest scale prototypes so far by utilizing the process of reticle stitching.

A full MOSS (or babyMOSS) prototype is made of three components: one so-called LEC (Left End-Cap), which supplies power and contains control and readout I/Os; ten (one, for babyMoss) RSUs (Repeated Sensor Units) which contain the pixel matrices; and one REC (Right End-Cap) which also supplies power. RSUs are split into top and bottom so-called `half-units', which each contain four pixel matrices. Matrices in the top half-units consist of $256~\times~256$ square \SI{22.5}{\micro\meter} pixels and $320~\times~320$ square \SI{18.0}{\micro\meter} pixels in the bottom half-units to test different pixel integration densities. The pixel sensing element is similar to DPTS, utilizing the n-gap layout, with two gap sizes implemented to study charge sharing for \SI{22.5}{\micro\meter} pixels. Four front-end variants were also implemented, but only the results from the so-called 'standard' (SF into folded-cascode) variant were presented. 

MOST has a similar LEC, RSU, REC structure but with an adjusted matrix size of $352\times62$, with four matrices per RSU, and utilizes \SI{18}{\square\micro\meter} pixels. A main difference compared to MOSS is the so-called granular powering scheme that allows small sections of the matrices to be selectively disconnected. This enables the isolation of local shorts or defects so they do not compromise the operation of the entire sensor. Some preliminary results from the characterization of MOSS and MOST will be included where applicable.

\begin{landscape}
%\begin{tiny}
    \begin{longtblr}[
                     caption = {Overview of prototypes in the TPSCo \SI{65}{\nano\meter} ISC technology.},
                     label = {tab:overview},
                     remark{Remarks} ={*Results of the characterization are not public at the time of writing.
                     $^{\dagger}$One of four matrices in the so-called \textit{top half-unit} in one so-called Repeated Sensor Unit (RSU).
                     $^{\ddagger}$One of four matrices in the so-called \textit{bottom half-unit} in one RSU.
                     $^{\S}$One of four submatrices in one RSU.
                     $^{\#}$Referred to as staggered in~\cite{ce65v2_pixel} and hexagonal in~\cite{ce65v2_iworid}. 
                     }
                    ]{colsep=3 pt,
                      row{1} = {font=\bfseries, c, guard},
                      row{2} = {guard},
                      %rowhead=2
                     }
    \toprule
    \SetCell[c=1]{c} Prototype   
        &  \SetCell[c=1]{l} Matrix
        &  \SetCell[c=1]{l} Pitch [\SI{}{\micro\meter}]
        &  \SetCell[c=1]{l} Sensor layout
        &  \SetCell[c=1]{l} Variants   
        &  \SetCell[c=3]{l} In-pixel front end
        &   &   &  \SetCell[c=1]{l} Output 
        & \\
% table body
    \cmidrule[r]{6-8}
    & (read out) & & & & Amplification & Discrimination & Digitization & \\
    \midrule[\lightrulewidth]
    \textbf{APTS}
    & \SetCell[r=4]{l} 4$\times$4 & 10$\times$10 & standard  & reference      & \SetCell[r=4]{l} SF with/without op amp & \SetCell[r=4]{l} no & \SetCell[r=4]{l} no & \SetCell[r=4]{l} analog \\
    &                             & 15$\times$15 & n-blanket & larger n-well  & & & & \\
    &                             & 20$\times$20 & n-gap     & smaller p-well & & & & \\
    &                             & 25$\times$25 &           & finger p-well  & & & & \\
\midrule[\lightrulewidth]
    \textbf{DPTS}
    & \SetCell[r=2]{l} 32$\times$32 & \SetCell[r=2]{l} 15$\times$15 & \SetCell[r=2]{l} n-gap & \SetCell[r=2]{l} none & high-gain cascoded & \SetCell[r=2]{l} yes & \SetCell[r=2]{l} yes & ToA \\
    &                               &                               &                        &                       & inverting amp.     &                      &                     & ToT  \\
\midrule[\lightrulewidth]
    \textbf{CE65}
    & \SetCell[r=3]{l} 64$\times$32 & 15$\times$15  & standard & \SetCell[r=3]{l} none & AC-coupled amp. & \SetCell[r=3]{l} no & \SetCell[r=3]{l} no & \SetCell[r=3]{l} analog \\
    &                               & 25$\times$25* & n-gap    &                       & DC-coupled amp. &                     &                     & \\
    &                               &               &          &                       & SF &                     &                     & \\
\midrule[\lightrulewidth]
    \textbf{CE65v2}
    & \SetCell[r=3]{l} 48$\times$24 & 15$\times$15     & standard  & square and  & \SetCell[r=3]{l} AC-coupled amp. & \SetCell[r=3]{l} no   & \SetCell[r=3]{l} no  & \SetCell[r=3]{l} analog \\
    &                               & 18$\times$18*    & n-blanket & staggered$^{\#}$  &                                  &                       &                      & \\
    &                               & 22.5$\times$22.5 & n-gap     &             &                                  &                       &                      & \\
\midrule[\lightrulewidth]
    \textbf{DESY-MLR1}
    & 2$\times$2 & 16.3$\times$16.3 & n-gap & none & CSA with Krumm.-type feedback & no & no  & analog \\
\midrule[\lightrulewidth]
    \textbf{DESY-ER1}
    & \SetCell[r=2]{l} 2$\times$2 & \SetCell[r=2]{l} 35$\times$25 & \SetCell[r=2]{l} n-gap & 2.5 and \SI{4.0}{\micro\meter} & \SetCell[r=2]{l} CSA with Krumm.-type feedback & \SetCell[r=2]{l} with/without & \SetCell[r=2]{l} no & analog \\
    & & & & gap width & & & & ToT, ToA \\
\midrule[\lightrulewidth]
    \textbf{H2M}
    & \SetCell[r=4]{l} 64$\times$16 & \SetCell[r=4]{l} 35$\times$35 & \SetCell[r=4]{l} n-gap & \SetCell[r=4]{l} none & \SetCell[r=4]{l} CSA with Krumm.-type feedback & \SetCell[r=4]{l} yes  & \SetCell[r=4]{l} yes & ToT (8-bit), \\
    & & & & & & & & ToA (8-bit), \\
    & & & & & & & & counting, \\
    & & & & & & & & triggered \\
\midrule[\lightrulewidth]
    \textbf{MOSS}
    & \SetCell[r=2]{l} 256$\times$256$^{\dagger}$ & \SetCell[r=2]{l} 22.5$\times$22.5 & \SetCell[r=4]{l} n-gap & 2.5 and \SI{5.0}{\micro\meter} & SF into folded-cascode & \SetCell[r=4]{l} yes & \SetCell[r=4]{l} yes & \SetCell[r=4]{l} ToT \\
    & mat & pix & lay & gap width & larger discriminating transistor & dis & dig & out \\
    & \SetCell[r=2]{l} 320$\times$320$^{\ddagger}$ & \SetCell[r=2]{l} 18.0$\times$18.0 & lay & \SetCell[r=2]{l} none & larger input transistor & dis & dig & out \\
    & mat & pix & lay & var & larger amplifying transistor & dis & dig & out \\
\midrule[\lightrulewidth]
    \textbf{MOST}
    &  \SetCell[r=2]{l} 352$\times$64$^{\S}$ & \SetCell[r=2]{l} 18$\times$18 & \SetCell[r=2]{l} n-gap & 2.5 and \SI{5.0}{\micro\meter} & \SetCell[r=2]{l} - & \SetCell[r=2]{l} yes & \SetCell[r=2]{l} yes & \SetCell[r=2]{l} ToT \\
      & mat & pix & lay & gap width & - & dis & dig & out \\
    \bottomrule
    \end{longtblr}
%\end{tiny}
\end{landscape}

\section{Experimental results}
\label{results}
Extensive measurement campaigns were conducted on the prototypes introduced in Section~\ref{prototypes}. The results allow for an assessment of the performance achievable in the TPSCo \SI{65}{\nano\meter} ISC technology and the trade-offs to be made between competing designs. They are summarized in this section, starting with general properties, which can usually be measured in the laboratory. A focus is put on the performance when it comes to charged-particle detection, and finally, the degradation of this performance after irradiation is addressed.

\subsection{General properties}
\label{results:general}
Some of the key properties of silicon sensors can be measured in the laboratory, for example using noise, test pulses, or radioactive sources like strontium-90 ($^{90}$Sr) or iron-55 ($^{55}$Fe). For details on the methodology, the reader is referred to the cited publications, while the results are summarized below.

\subsubsection{Pixel capacitance}
\label{results:general:cap}
As outlined in~\cite{snoeys_cap}, the ratio of the collected charge over the input capacitance of the in-pixel amplifier $Q / C$ needs to be as high as possible to limit the analog power consumption of the in-pixel front end. This means that not only the charge collection but also the pixel capacitance---a significant contribution to the input capacitance---needs to be optimized. The SF employed in the in-pixel front end of the APTS allows measuring the pixel capacitance, and a systematic study of different chip flavors is presented in~\cite{apts}. It is observed that the pixel capacitance is independent of the pitch but reduces with the reverse bias voltage as the depletion zone grows. For the n-gap and n-blanket layout, it reduces from \SI{3}{\femto\farad} to \SI{2}{\femto\farad} between \SI{-1.2}{\volt} and \SI{-4.8}{\volt}\footnote{In this case, the bulk depletion changes very little with the reverse bias voltage. The change in capacitance is caused by the depletion of the volume close to the collection electrode, causing the effective size of the electrode to shrink.}. The dependence is less pronounced for the standard layout, and the capacitance is close to \SI{2}{\femto\farad} in the above-given voltage range. Similar measurements on APTS with operational-amplifier readout are presented in~\cite{dis_russo} and show the same behavior.

\subsubsection{Noise, threshold dispersion, and fake-hit rate}
\label{results:general:noise}
\begin{figure}[tbp]
    \centering
    \includegraphics[width=1\linewidth]{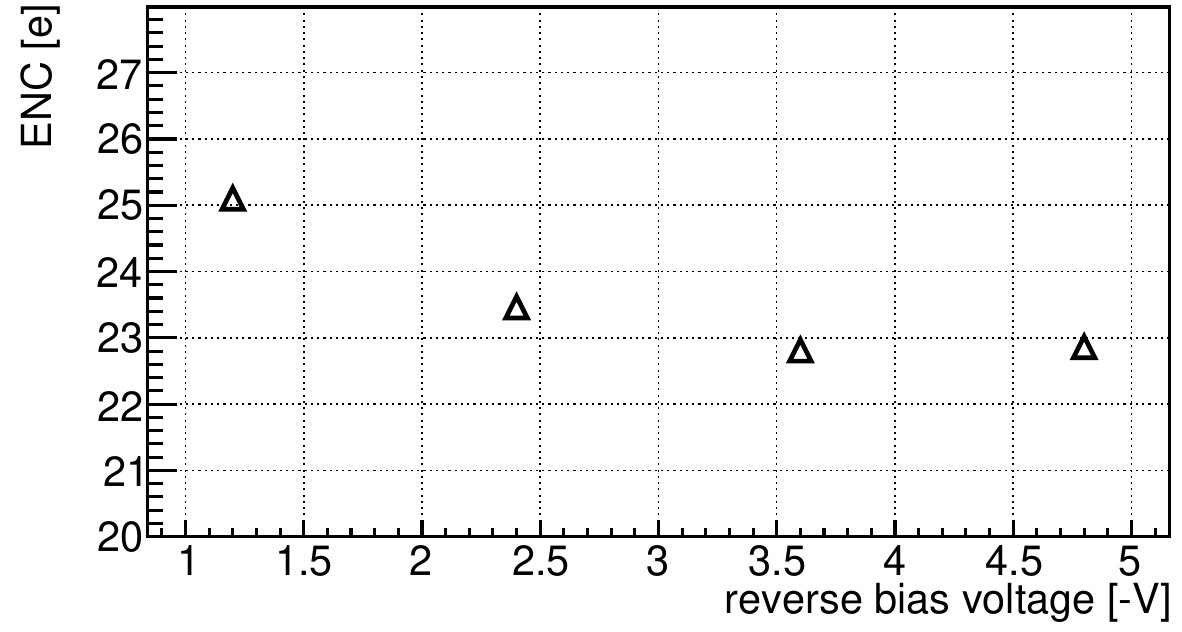}
    \caption{ENC for APTS with operational-amplifier readout, a pitch of \SI{10}{\micro\meter}, and n-gap layout as a function of the reverse bias voltage. The numbers are from~\cite{apts-timing} and derived as the RMS of the signal baseline. Statistical uncertainties are below $0.05\,\text{e}^{-}$ and not visible in the figure.}
    \label{fig:noise_voltage}
\end{figure}

In this context, noise is to be understood as the fluctuations around the baseline signal in the absence of a particle interaction. For many of the prototypes, the noise is measured in the laboratory, and expressed in terms of an equivalent noise charge (ENC). It depends on the type and operation point of the front end, and reported values range from $23\,\text{e}^{-}$ to $36\,\text{e}^{-}$ for APTS~\cite{apts,apts-timing,apts_desy_sim}, $15\,\text{e}^{-}$ to $25\,\text{e}^{-}$ for CE65~\cite{ce65_vci}, down to $9\,\text{e}^{-}$ for DPTS at a reverse bias voltage of \SI{-1.2}{\volt}~\cite{dpts}, and $15\,\text{e}^{-}$ for MOSS~\cite{moss} and MOST~\cite{most}. Figure~\ref{fig:noise_voltage} shows the dependence on the bias voltage, which is assumed to be dominated by the above-mentioned change in capacitance. The same behavior is observed in~\cite{apts} and~\cite{dis_russo}, including also measurements on sensors with n-blanket and standard layout. The noise for sensors in the standard layout is reported to be systematically lower and less dependent on the reverse bias voltage, as expected from the capacitance measurements described above. One has to note that changes of the reverse bias voltage might also affect the operation point of the front-end electronics, as it is applied to the substrate and the p-well.

A key property of a sensor is the detection threshold, which is usually expressed as the minimal charge required to trigger the readout of an individual pixel. For the considered type of sensors, typical operational thresholds are in the range between $100\,\text{e}^{-}$ and $200\,\text{e}^{-}$. It requires finding a compromise between high hit-detection efficiency and the rate of fake hits originating from noise. The constraint on this fake-hit rate is typically given by the experiment and defines a minimal threshold based on an occupancy measurement in a dark environment. For sensors with more than a few pixels, this depends also on the threshold dispersion, which characterizes the variation of the per-pixel detection threshold with respect to a global threshold level.

The threshold dispersion of DPTS depends on the operation point of the front end. The smallest reported threshold dispersion is $13\,\text{e}^{-}$, which is larger than the noise contribution of $9\,\text{e}^{-}$ for the same setting~\cite{dpts}. The threshold dispersion of the MOSS is reported to be $18\,\text{e}^{-}$, compared to the ENC of $15\,\text{e}^{-}$, for a reference region with \SI{22.5}{\micro\meter} pitch and a gap size of \SI{2.5}{\micro\meter}~\cite{moss}. For H2M, a threshold dispersion on the order of $25\,\text{e}^{-}$ and an ENC on the order of $40\,\text{e}^{-}$ are reported~\cite{h2m}.

Fake-hit rates are subject to great variations between prototypes and depend strongly on the operation point of the in-pixel front end and the threshold in particular. Detailed studies on DPTS are presented in~\cite{dpts2}. It is shown that the fake-hit rate is below \SI{1e-2}{\hertz} per pixel for thresholds above $100\,\text{e}^{-}$ and most operation points of the in-pixel front end. It is also demonstrated that this can be reduced by more than two orders of magnitude by masking the noisiest \SI{1}{\percent} of the pixels at this threshold. Measurements on MOSS show larger fake-hit rates on the order of \SI{1}{\hertz} per pixel at this threshold, which might be due to a parasitic capacitive coupling between analog bias and strobe-distribution lines~\cite{moss}. For H2M, a fake-hit rate of \SI{18.4}{\kilo\hertz} per pixel at a threshold of $144\,\text{e}^{-}$ is reported~\cite{h2m}. The reason for this comparably large fake-hit rate, is assumed to be the larger noise and threshold dispersion quoted above.

\subsubsection{Power dissipation}
\label{results:general:power}
In many applications, the power dissipated by the analog and digital in-pixel front end and the periphery of a pixel sensor needs to stay below stringent limits, to constrain the extent of cooling infrastructure in order to minimize multiple Coulomb scattering~\cite{moliere,bethe} in the detector volume. However, systematic studies on the power dissipation are scarce, as many prototype families are in an early stage of development. The power consumption of the DPTS matrix depends on the operation point of the front end, and ranges from \SI{12}{\nano\watt} to \SI{120}{\nano\watt} per pixel, corresponding to a matrix power consumption between \SI{5.3}{\milli\watt\per\square\centi\meter} and \SI{53}{\milli\watt\per\square\centi\meter}~\cite{dpts2}. The results in~\cite{dpts} were obtained for the operating point with the largest power consumption, while the effect of reducing the power consumption is presented in~\cite{dpts2}. For MOSS, the analog power density is reported to be \SI{7}{\milli\watt\per\square\centi\meter} and \SI{11}{\milli\watt\per\square\centi\meter} for top and bottom half-units, respectively~\cite{moss_proceeding}. These two regions feature different pixel densities, employing pixel pitches of \SI{18}{\micro\meter} and \SI{22.5}{\micro\meter}. One should note that the analog front end of DPTS and MOSS are both based on the design presented in~\cite{DPTS_FE}. %For APTS with operational-amplifier readout, the power consumption amounts to \SI{150}{\micro\watt} per in-pixel front end~\cite{apts-timing}.

\subsection{Charged-particle detection}
\label{results:mip}
The key objective of a pixel sensor for tracking or vertexing applications is the detection of relativistic charged particles, and in the worst case, said particles are MIPs. The discussed prototypes have been tested at different facilities, e.g. the DESY II Test Beam Facility~\cite{desyII} and CERN PS or SPS~\cite{SPS}, using different reference systems, like an EUDET-type~\cite{eudet_tel} or an ALPIDE-based~\cite{alpide2016} beam telescope. The key observables quantifying the MIP detection capabilities of a silicon sensor are: the probability to detect an impinging MIP, the hit-detection efficiency; the accuracy of the impact-position measurement in the sensor plane, the spatial hit resolution; and the accuracy of the measurement of the interaction time, the temporal hit resolution. These properties depend on environmental and operational conditions and will be discussed in the following sections. In all cases the incidence of the traversing MIPs is orthogonal to the sensor surface.

\subsubsection{Charge collection and charge sharing}
\label{results:mip:q}
Even though all the investigated prototypes have been produced on the same, or very similar, base material, their charge collection properties may be different. This is because the size and shape of the depletion region and the properties of the electric field within depend on prototype characteristics, like pitch and sensor layout, and operation conditions, like the reverse bias voltage. Within the depletion region, charge transport is dominated by drift towards the collection electrode. Outside the depletion region, charge transport is governed by diffusion. Diffusion is the main cause for charge sharing---the effect that a single particle interaction might induce a signal in several neighboring pixels. Neighboring pixels with a signal are grouped into so-called clusters, and the number of pixels in a cluster is referred to as cluster size.
\begin{figure}[tbp]
    \centering
    \includegraphics[width=1\linewidth]{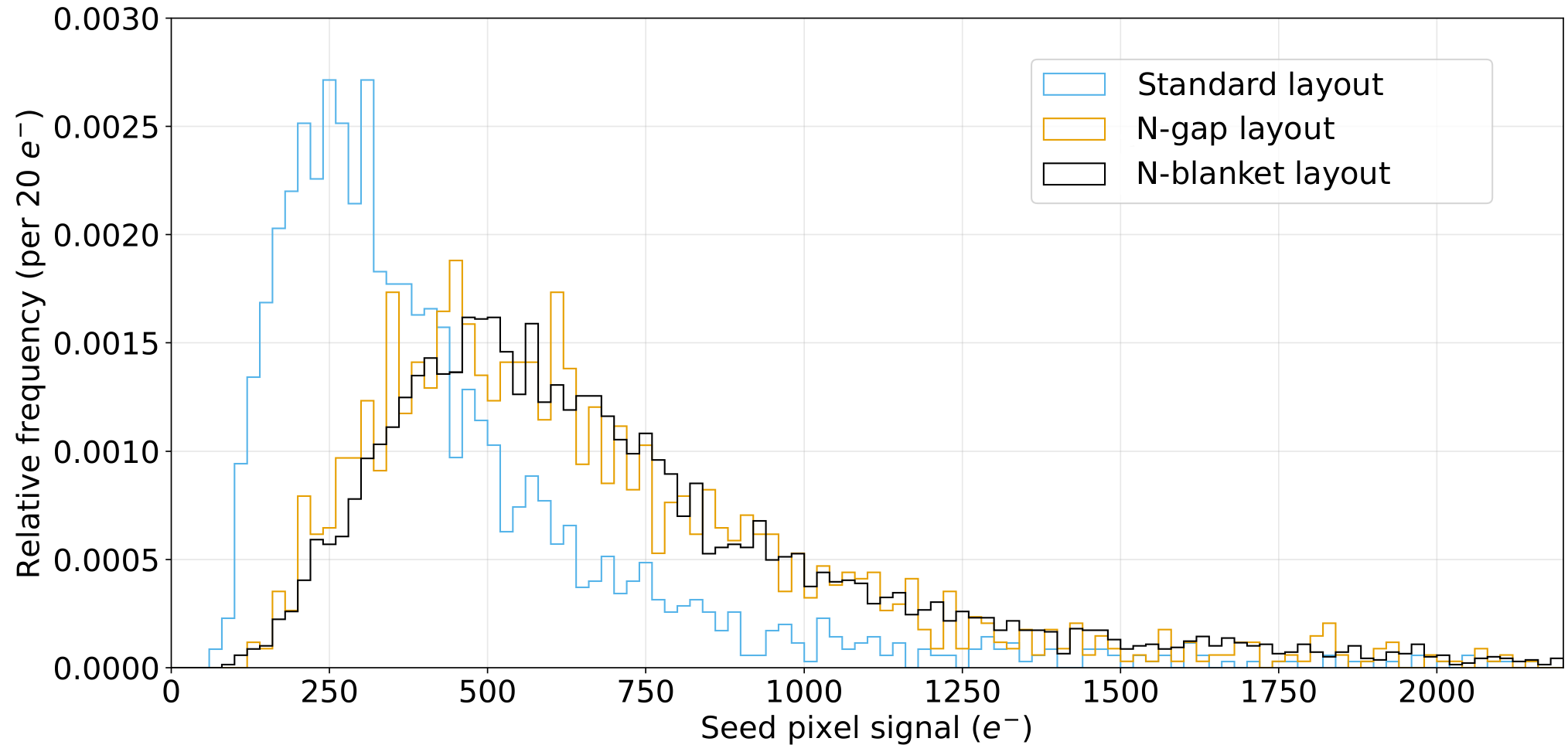}
    \caption{Distributions of the seed pixel signal for different types of APTS sensors. All sensors have a pitch of \SI{15}{\micro\meter}, and are reverse biased to \SI{-1.2}{\volt}. Modified from~\cite{apts}.}
    \label{fig:landau}
\end{figure}

Figure~\ref{fig:landau} shows the distribution of the seed pixel signal, defined as the largest pixel charge in a cluster, for APTS with different sensor layouts. The data have been recorded at CERN SPS and shows that the charge collection for the n-blanket and n-gap layouts are similar, with a most probable value of $500\,\text{e}^{-}$~\cite{apts}. Similar measurements on the H2M prototype yield a most probable value of $680 \pm 20\,\text{e}^{-}$, and the differences to the APTS results are attributed to the calibration procedure~\cite{h2m}. Calculating the expected signal in \SI{10}{\micro\meter} of silicon, using a parameterization from~\cite{bichsel} and a mean ionization energy of \SI{3.66}{\electronvolt}~\cite{scholze}, results in approximately $620\,\text{e}^{-}$. A simulation using the Allpix$^2$~\cite{Allpix2} framework, in conjunction with Geant4~\cite{Geant4_1, Geant4_2, geant4_3}, employing the Photo Absorption Ionization model~\cite{pai}, yields a deposited charge of about $550\,\text{e}^{-}$ in \SI{10}{\micro\meter} of silicon for \SI{120}{\giga\electronvolt} pions. This motivates the following interpretation: the sensitive region is given by the epitaxially grown high-resistivity layer, reduced due to dopant diffusion from the substrate during growth of the epitaxial layer, and structures required for the integrated circuitry. The high doping concentration in the above-mentioned regions reduces the carrier lifetime, such that electron-hole pairs generated in this region recombine to a large part. The thickness of the epitaxial layer is reported to be \SI{10}{\micro\meter} for the discussed TPSCo \SI{65}{\nano\meter} ISC technology~\cite{Snoeys:2023hF}, which explains the APTS results below both $620\,\text{e}^{-}$ and $550\,\text{e}^{-}$. This explains why thinning down to a total thickness of \SI{21}{\micro\meter} is possible, without significant effects on key performance characteristics of the sensor, as demonstrated in~\cite{h2m}. This leaves around \SI{5}{\micro\meter} for metal-interconnection layers~\cite{h2m} and a few micrometer of low-resistivity substrate for bias-voltage distribution. Typically, the investigated prototypes have a total physical thickness around~\SI{50}{\micro\meter}~\cite{moss, h2m}.

Figure~\ref{fig:landau} includes also the seed pixel signal distribution for the standard layout, which peaks at a smaller value. This is due to the partially diffusion-driven charge transport leading to more charge sharing, hence a larger average cluster size for this layout. Since a cluster size larger than one allows to interpolate the hit position between the pixel centers, charge sharing has a beneficial effect on the spatial resolution (see Section~\ref{results:mip:res_spat}). On the other hand, the smaller seed pixel signal has a detrimental effect on the hit-detection efficiency (see Section~\ref{results:mip:eff}), and a larger cluster size also means that more hits need to be read out.
\begin{figure}[tbp]
    \centering
    \includegraphics[width=1\linewidth]{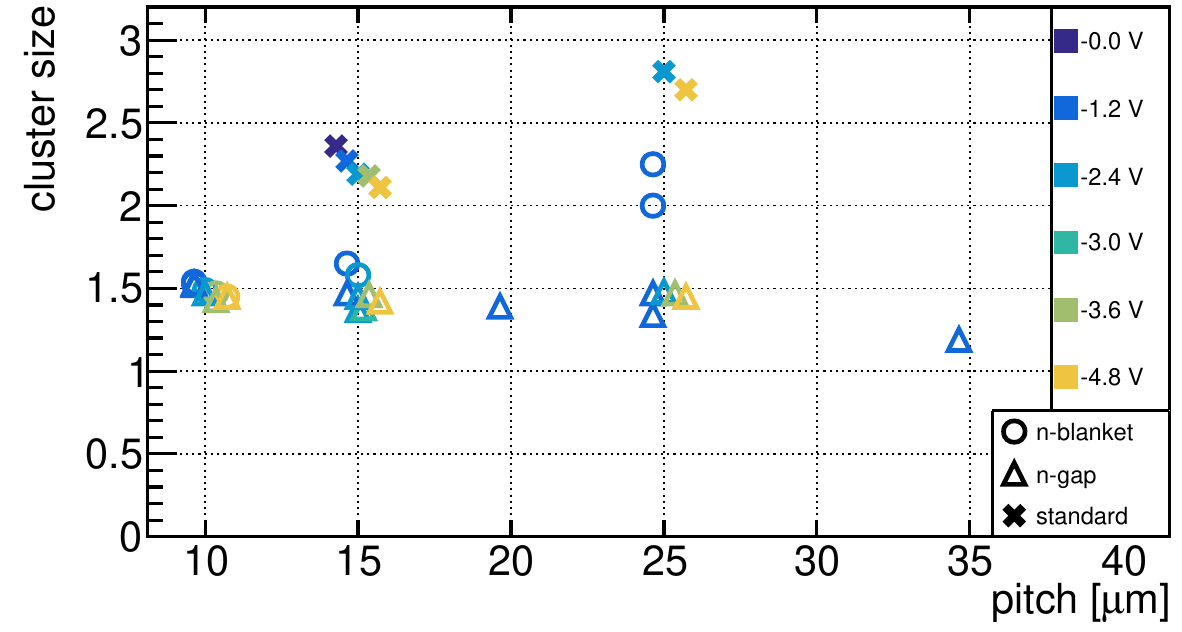}
    \caption{Cluster size at a threshold of $100\,\text{e}^{-}$, comparing different pixel pitches, reverse bias voltages, and sensor layouts. All pitches are multiples of \SI{5}{\micro\meter}, the points are shifted in x-direction for better visibility. The results represent various prototypes in the TPSCo \SI{65}{\nano\meter} ISC technology and are taken from~\cite{apts,apts-timing,dpts,h2m,apts_desy_sim,dis_adri}. A detailed listing of the numbers and the corresponding references is given in \ref{app:tab}.}
    \label{fig:clustersize}
\end{figure}

A systematic comparison of cluster sizes for different sensor layouts, pitches, and reverse bias voltages is presented in Figure~\ref{fig:clustersize}. For the standard and n-blanket layout, it shows an increase of the cluster size as a function of the pitch, because the contribution of charge transport by diffusion increases. For the n-gap layout, the lateral electric field at the pixel boundary is given by the doping gradient and width of the gap, and hence not a strong function of the pitch. Thus, the size of the region in which charge sharing takes place decreases relatively as the pitch increases. A systematic study of the dependence of the cluster size on the reverse bias voltage is presented in~\cite{apts}, and observed to be smaller than \SI{10}{\percent} per volt.

In general, one should consider, that collected charge and cluster size depend on the response function (effective integration time) of the front end, which differs between prototypes and operation conditions. This might have a subtle impact on hit-detection efficiency and also spatial resolution.

\subsubsection{Hit-detection efficiency}
\label{results:mip:eff}
The hit-detection efficiency of the investigated prototypes depends on the pitch, and sensor layout (standard, n-blanket, or n-gap) of the studied sample, as well as operation parameters like detection threshold, and the reverse bias voltage applied to p-well and substrate.
\begin{figure}[tbp]
    \centering
    \includegraphics[width=1\linewidth]{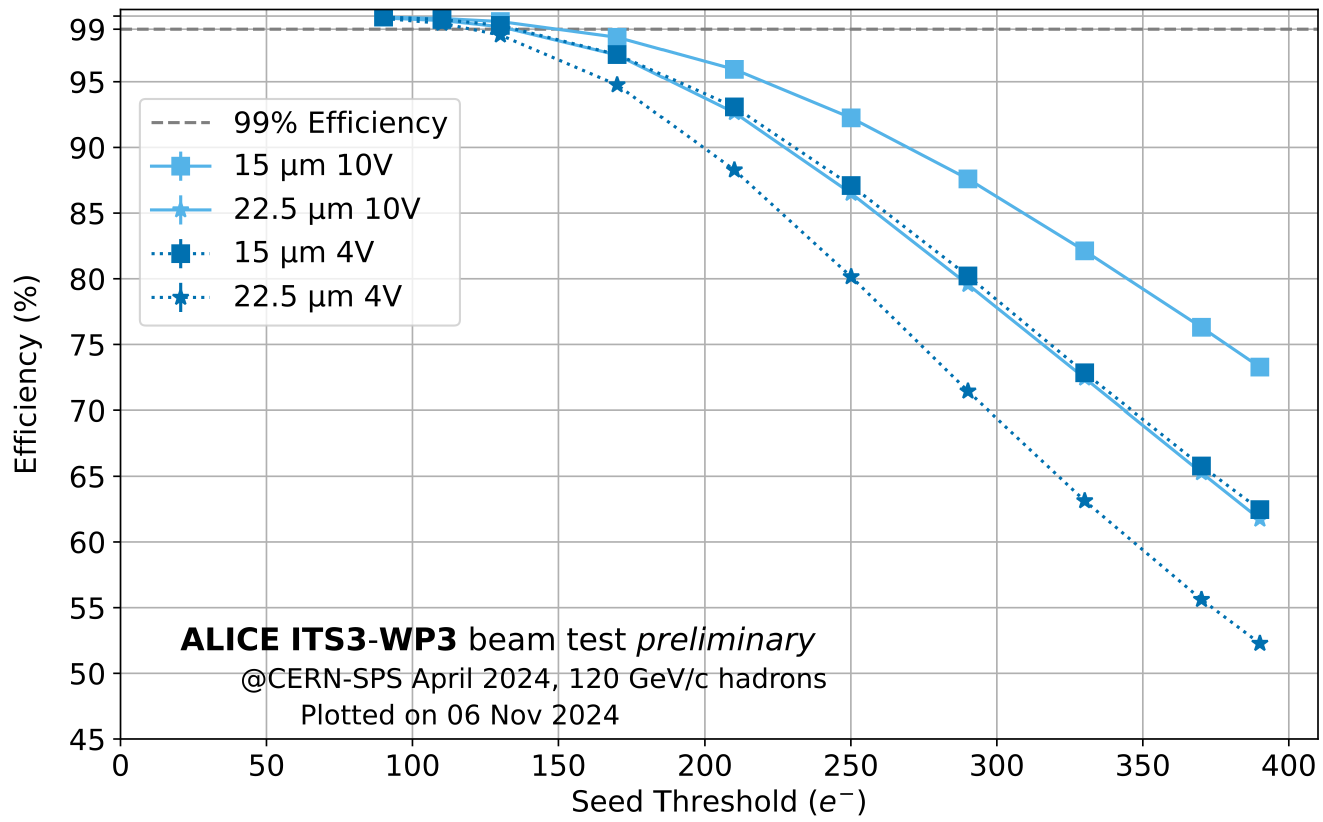}
    \caption{Hit-detection efficiency as a function of the threshold, comparing two CE65v2 chips in standard layout with different pitches and at different reverse bias voltages (absolute values). Modified from~\cite{ce65v2_pixel}.}
    \label{fig:eff_pitch}
\end{figure}

The dependence of the hit-detection efficiency on the threshold can be used to assess the operational margin of an investigated prototype. Figure~\ref{fig:eff_pitch} shows this threshold dependency for two flavors of the CE65v2 chip with pitches of \SI{15}{\micro\meter} and \SI{22.5}{\micro\meter}, both standard layout, operated at reverse bias voltages of \SI{4}{\volt} and \SI{10}{\volt}. It shows that the efficiency drops continuously in both cases, and faster for the larger pitch and smaller reverse bias voltage. To make it easier to compare different samples and operation conditions, the value $T_{99}$ is used, defined as the threshold where the efficiency reaches \SI{99}{\percent} (a common choice in tracking applications~\cite{apts,cms_p2}).

\begin{figure}[tbp]
    \centering
    \includegraphics[width=1\linewidth]{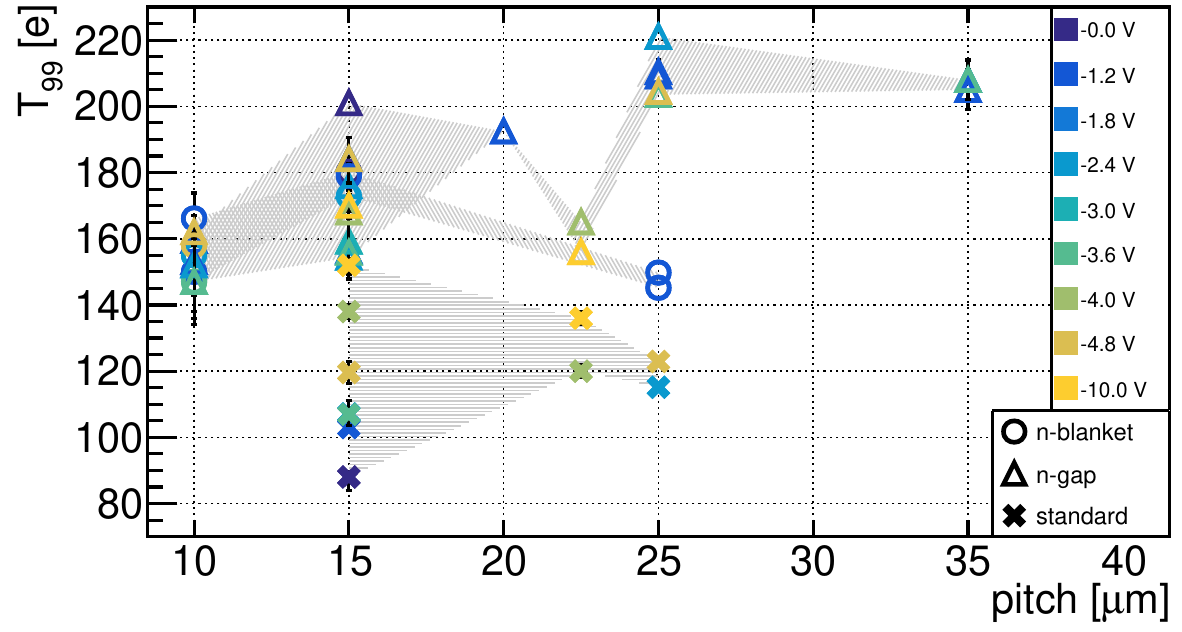}
    \caption{Threshold requirement for \SI{99}{\percent} efficiency for different pixel pitches, reverse bias voltages, and sensor layouts. The gray areas indicate the range covered by the three layouts. The results represent various prototypes in the TPSCo \SI{65}{\nano\meter} ISC technology and are taken from~\cite{apts,apts-timing,dpts,ce65v2_pixel,h2m,apts_desy_sim,dis_adri}. A detailed listing of the numbers and the corresponding references is given in \ref{app:tab}.}
    \label{fig:eff_mega}
\end{figure}

Figure~\ref{fig:eff_mega} summarizes the performance of various prototypes produced in the TPSCo \SI{65}{\nano\meter} ISC technology by presenting $T_{99}$ as a function of the pixel pitch, for all three sensor layouts and available reverse bias voltages. In general, the n-gap layout reaches the highest $T_{99}$, ranging from $147\,\text{e}^{-}$ to $221\,\text{e}^{-}$, while n-blanket ranges from $145\,\text{e}^{-}$ to $180\,\text{e}^{-}$, and standard from $88\,\text{e}^{-}$ to $152\,\text{e}^{-}$. Considering typical ENCs on the order of $30\,\text{e}^{-}$, these higher $T_{99}$ values represent a significant advantage of the n-gap and n-blanket layout. 

It is not possible to make further conclusions on the dependence of $T_{99}$ on reverse bias voltage and pitch based on this figure, as such trends are overshadowed by a combination of systematic differences between the measurements and prototype properties\footnote{For example, systematic uncertainties on the charge calibration, which are not always discussed in the references, different selection criteria, or charge integration times.}. On the other hand, the data presented in~\cite{apts} indicates an increase of $T_{99}$ with decreasing reverse bias voltage for standard layout prototypes with a pitch of \SI{15}{\micro\meter}. This is confirmed in~\cite{ce65v2_pixel}, for pitches of \SI{15}{\micro\meter} and \SI{22.5}{\micro\meter}. Such a clear trend is observed for neither n-blanket nor n-gap sensors, supposedly because the depleted volume hardly changes in the considered range. One should note that the impact of the reverse bias voltage is driven by a complex interplay between the growth of the depleted volume with a certain impact on capacitance and charge collection and shifting of the NMOS transistor operation points within the deep p-well. As the former effect is potentially beneficial, especially for the standard layout, one should keep it in mind for further generations of sensors and consider the benefits of AC-coupled readout electronics. 
The results presented in~\cite{apts} also indicate, that an increasing pitch (shown between \SI{10}{\micro\meter} and \SI{25}{\micro\meter}) has a positive impact on the efficiency of the n-gap layout, presumably, because there is less charge sharing for geometric reasons (see also Section~\ref{results:mip:q}). This trend is reproduced in~\cite{ce65v2_pixel}, where pitches of \SI{15.0}{\micro\meter} and \SI{22.5}{\micro\meter} are compared, although the corresponding $T_{99}$ in Figure~\ref{fig:eff_mega} show a different behavior. This is probably caused by the small slope of the corresponding curves, which makes $T_{99}$ sensitive to sample-to-sample variations.

In general, when comparing different sensor layouts, one can assume that the reduction of charge sharing for geometric reasons is at some point countered by a reduction of the lateral electric field, which impairs charge collection. The deterioration of the charge collection with increasing pitch is expected to happen latest for the n-gap, earlier for the n-blanket, and even earlier for the standard layout. One interesting example is the H2M chip with n-gap layout and a pitch of \SI{35}{\micro\meter}. The lateral electric field is small and distorted through the presence of large n-wells within the deep p-well (see Figure~\ref{fig:layout}), which slows the charge collection and leads to a non-uniform in-pixel efficiency due to the fast front end~\cite{h2m,h2m_meas}. The simulations presented in~\cite{h2m,h2m_sim} reproduce this effect.

\subsubsection{Spatial hit resolution}
\label{results:mip:res_spat}
The spatial hit resolution of a pixel sensor, for particle tracking or vertex reconstruction, is usually derived from the width of a residual distribution. The residual is calculated as the difference between the intercept of the reconstructed track with the sensor plane and the sensor's position estimate. The residuals width needs to be corrected for the tracking uncertainty to infer the intrinsic spatial hit resolution of the investigated sensor. The sensor's position estimate is calculated for groups of neighboring pixels defining a cluster, and typically the average of the pixel center positions. If the pixel response contains information about the collected charge, this average might be charge-weighted to interpolate between the pixel centers. As the distribution of the charge between two neighboring pixels is not necessarily a linear process, this charge-weighted average can be corrected for the introduced bias, usually referred to as $\eta$-correction~\cite{eta}. Also, the precision of the charge measurement, affected by calibration, a finite granularity (number of bits), and noise, has an impact on the accuracy of the interpolation. If the pixel response is binary (hit or no hit), such an interpolation is not possible.
\begin{figure}[tbp]
    \centering
    \includegraphics[width=1\linewidth]{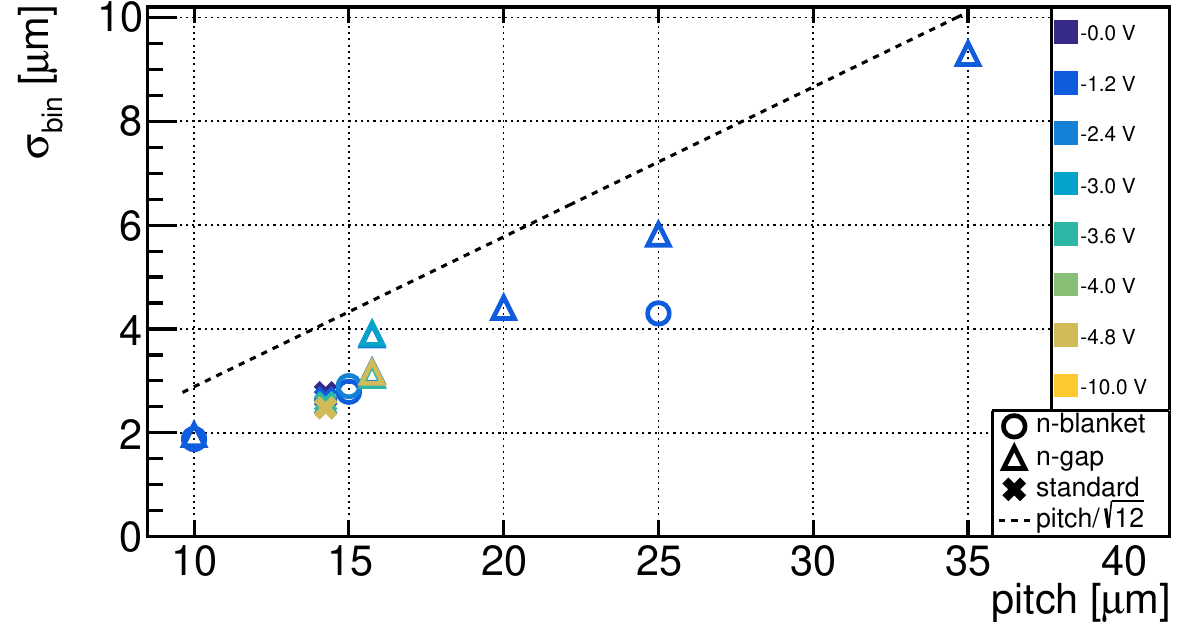}
    \caption{Spatial hit resolution without using charge information for different pixel pitches, reverse bias voltages, and sensor layouts, plotted for a detection threshold of $100\,\text{e}^{-}$. The points at a pitch of \SI{15}{\micro\meter} are slightly shifted in x-direction for better visibility. The results represent various prototypes in the TPSCo \SI{65}{\nano\meter} ISC technology and are taken from~\cite{apts, dpts, h2m}. A detailed listing of the numbers and the corresponding references is given in~\ref{app:tab}.}
    \label{fig:res_bin}
\end{figure}

\begin{figure}[tbp]
    \centering
    \includegraphics[width=1\linewidth]{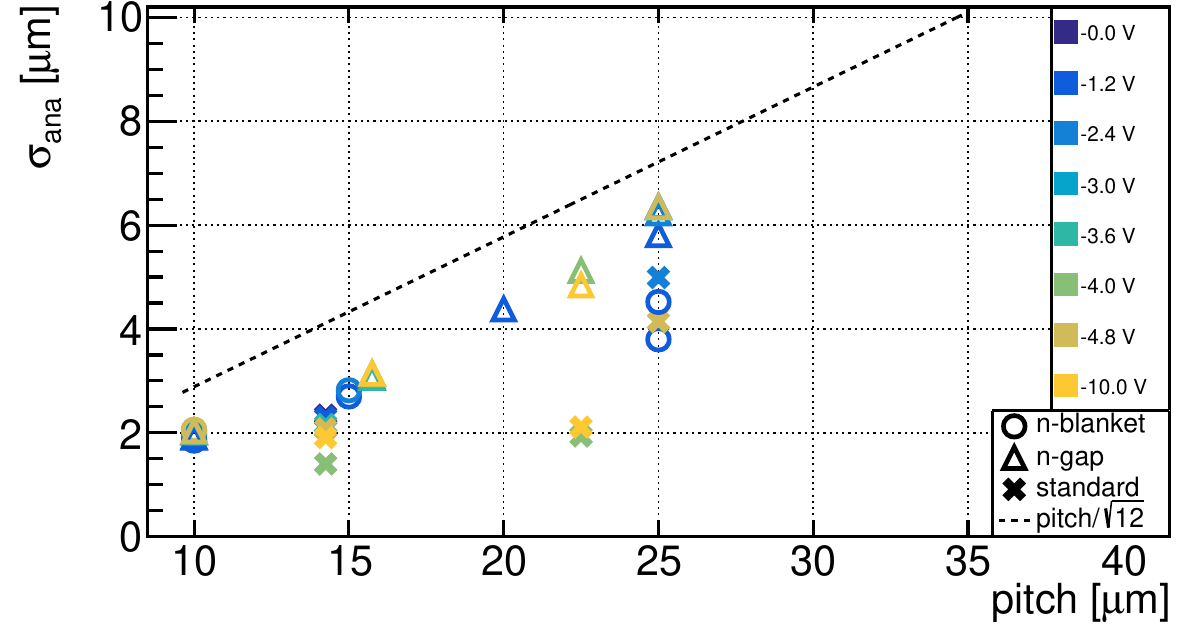}
    \caption{Spatial hit resolution using charge information (sometimes $\eta$-corrected) for different pixel pitches, reverse bias voltages, and sensor layouts. Plotted for a detection threshold of $100\,\text{e}^{-}$. The points at a pitch of \SI{15}{\micro\meter} are slightly shifted in x-direction for better visibility. The results represent various prototypes in the TPSCo \SI{65}{\nano\meter} ISC technology and are taken from~\cite{apts, apts-timing, ce65v2_pixel, apts_desy_sim, dis_adri}. A detailed listing of the numbers and the corresponding references is given in~\ref{app:tab}.}
    \label{fig:res_ana}
\end{figure}

Figure~\ref{fig:res_bin} shows the spatial hit resolution for a binary pixel response $\sigma_{\text{bin}}$, for various prototypes and reverse bias voltages as a function of the pitch. The measurements are $\mathcal{O}(\SI{1}{\micro\meter})$ better than expectation without charge sharing (pitch$/\sqrt{12}$). Considering also Figure~\ref{fig:clustersize}, this indicates the significance of the beneficial charge-sharing effect. The largest number of data points is available for a pitch of \SI{15}{\micro\meter}, where most results are closely grouped, but still indicating an improvement on the order of \SI{1}{\micro\meter} when moving from an n-gap layout to the standard layout. The outliers at a resolution of about \SI{4}{\micro\meter} were measured after irradiation~\cite{dpts} (see~\ref{app:tab} for details), which explains a small reduction in performance. The data points at \SI{25}{\micro\meter} show that the impact of charge-sharing effects becomes more relevant at this larger pitch. One can expect a similar improvement for the standard layout, but systematic studies are not publicly available at the time of writing.

Including charge information, as shown in Figure~\ref{fig:res_ana}, leads to similar resolution $\sigma_{\text{ana}}$ for the n-gap layout. On the other hand, the results for the n-blanket and standard layout are significantly better. Comparing the results for the n-blanket and the standard layout at a pitch of \SI{25}{\micro\meter} reveals some inconsistencies, which are likely caused by differences in the reference system, different methods to quantify the tracking uncertainty, the fact that not all references apply an $\eta$-correction, and the different methods used to determine the residual width. This complicates the comparison of spatial resolutions; further detail is given in~\ref{app:tab}. One should note that the effect of the charge-measurement granularity can be considered negligible for the presented measurements because it is either fine enough or has very limited impact because of the small cluster size.
\begin{figure}[tbp]
    \centering
    \includegraphics[width=1\linewidth]{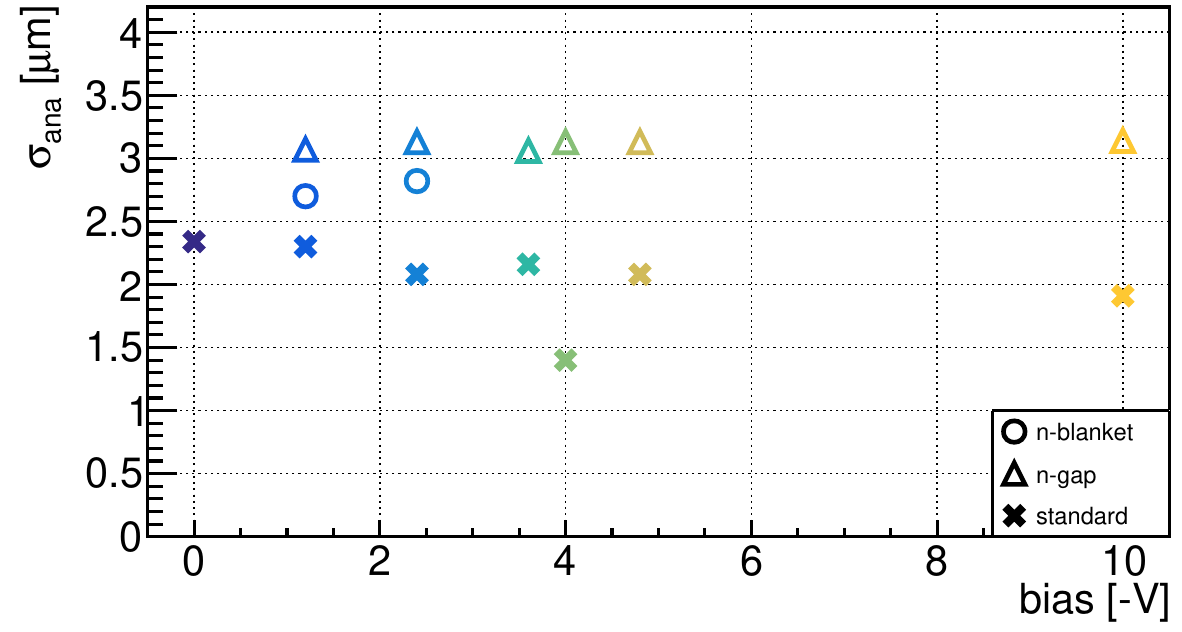}
    \caption{Spatial hit resolution using charge information (sometimes $\eta$-corrected) for prototypes with a pixel pitch of \SI{15}{\micro\meter}, as a function of the reverse bias voltages, showing different sensor layouts. plotted for a detection threshold of $100\,\text{e}^{-}$. The results represent various prototypes in the TPSCo \SI{65}{\nano\meter} ISC technology and are taken from~\cite{apts, dpts, ce65v2_pixel}. A detailed listing of the numbers and the corresponding references is given in~\ref{app:tab}.}
    \label{fig:res_volt}
\end{figure}

Figure~\ref{fig:res_volt} shows the dependence of the spatial hit resolution on the reverse bias voltage. The increasing reverse bias voltage is expected to have a beneficial effect on the charge collection, which is in turn beneficial for the spatial resolution. On the other hand, the increasing reverse bias voltage reduces charge sharing, as discussed in Section~\ref{results:mip:q}, which has a detrimental effect on spatial resolution. The overall effect is observed to be negligible for the n-gap layout, while the small number of data points does not allow conclusions about the behavior of the n-blanket layout. The results for the standard layout from~\cite{ce65v2_pixel} (-4 and \SI{-10}{\volt}) suggest that the spatial resolution deteriorates with increasing absolute bias, while the results from~\cite{apts}, suggest the opposite dependence. This may be explained by signal saturation, affecting the results at \SI{-10}{\volt} presented in~\cite{ce65v2_pixel}

\begin{figure}[tbp]
	\centering
	\includegraphics[width=1\linewidth]{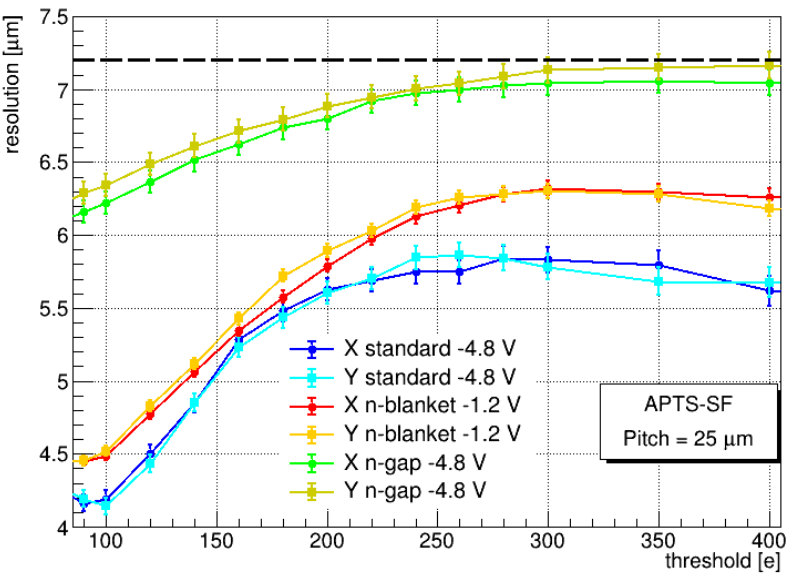}
	\caption{Spatial hit resolution using charge information as a function of the detection threshold, comparing different sensor layouts. The differences in bias voltage are considered to be negligible. Taken from~\cite{dis_adri}.}
	\label{fig:res_layout}
\end{figure}

Figure~\ref{fig:res_layout} shows the dependence of the spatial resolution on the detection threshold, including all three sensor layouts for a pitch of \SI{25}{\micro\meter}. For all three sensor layouts, the spatial resolution deteriorates with increasing thresholds up to $250\,\text{e}^{-}$, and much faster for the n-blanket and the standard layout. This renders the detection threshold a key parameter to optimize, as it has a significant impact on both hit-detection efficiency and spatial hit resolution. For thresholds above $250\,\text{e}^{-}$, the spatial resolution of the standard and n-blanket layout starts to improve with increasing threshold. This is an artifact caused by the fact that the efficiency reduction towards high thresholds affects the pixel borders first. For the n-gap layout this happens at even higher thresholds.

\subsubsection{Temporal hit resolution}
\label{results:mip:res_temp}
Like the spatial hit resolution, the temporal hit resolution of a detector for MIPs is determined with the aid of a reference system. After reconstructing and selecting tracks, the time stamps provided by the investigated and the reference detector are used to derive residuals. The shape of the residual distribution is determined by the temporal response of both detectors, so that the temporal response of the investigated sensor can be inferred if the reference system is well characterized or its contribution is negligible.

Literature on the temporal hit resolution of the covered \SI{65}{\nano\meter} prototypes is scarce compared to hit-detection efficiency and spatial hit resolution, and therefore a comprehensive overview over the full parameter space covered by the different prototypes (see Table~\ref{tab:overview}) cannot be given. Results on the characterization of APTS with operational-amplifier readout are presented in~\cite{apts-timing} and cover the n-blanket and n-gap layout at a pitch of \SI{10}{\micro\meter} and reverse bias voltages between \SI{1.2}{\volt} and \SI{4.8}{\volt}. The waveforms are recorded with an oscilloscope, and a temporal hit resolution of $\SI{82}{\pico\second} \pm \SI{3}{\pico\second}$, is reported for the highest (best) reverse bias voltage and independent of the sensor layout. These results are achieved by applying a constant-fraction discrimination. Nevertheless, variations of the measured arrival time as a function of the in-pixel position are reported. These are correlated with the cluster size, so that the cluster size can be used to derive a correction. After applying this correction, the temporal resolution for the n-gap layout is reported to reach $\SI{63}{\pico\second} \pm \SI{3}{\pico\second}$ at the highest reverse bias voltage. Characterization in a setup with a $^{90}$Sr radioactive source~\cite{dis_russo} show comparable results for the n-gap and n-blanket layout. A temporal hit resolution of $\SI{189}{\pico\second} \pm \SI{2}{\pico\second}$ is reported for the standard layout\footnote{This is after subtracting the time resolution of the reference system of about \SI{40}{\pico\second}, which is mentioned in reference~\cite{dis_russo} but not subtracted in the presented figures. The other results are quoted as given in the reference, after subtraction of the reference system's contribution. The results quoted in this paragraph are derived as the RMS of the central \SI{99.7}{\percent} of the entries in the corresponding residual distributions.}.

Variations of the arrival time as a function of the in-pixel position increase with the pixel size. They are on the order of \SI{60}{\nano\second} in~\cite{h2m_meas} for H2M with a pitch of \SI{35}{\micro\meter} and n-gap layout, as opposed to \SI{120}{\pico\second} for APTS with a pitch of \SI{10}{\micro\meter}~\cite{apts-timing}. This explains the larger time resolution reported for H2M, namely  $\SI{28.4}{\nano\second} \pm \SI{0.2}{\nano\second}$, for a reverse bias voltage of \SI{-3.6}{\volt}~\cite{h2m}. For DESY-ER1, with a pitch of $25 \times \SI{35}{\square\micro\meter}$, operated at a reverse bias voltage of \SI{-3.6}{\volt}, the time resolution is constrained to be better than \SI{8.72}{\nano\second}~\cite{desy_er1}. In general, it can be assumed that the effect of the pitch on the time resolution will be more severe for the n-blanket and standard layout.

For DTPS, with a pitch of \SI{15}{\micro\meter} and n-gap layout, a time resolution of $\SI{6.3}{\nano\second} \pm \SI{0.1}{\nano\second}$ is reported~\cite{dpts}, employing ToT-based time-walk correction. It is expected that these results can be improved by increasing the front-end current, hence the power consumption. This implies that the reported results are limited by the front end rather than the charge collection process. This is in line with results from DESY-MLR1~\cite{desy-mlr1}, where rise times between \SI{4}{\nano\second} and \SI{9}{\nano\second} are reported and assumed to be limited by the last amplifier stage.

\subsection{Radiation hardness}
\label{reslts:rad}
The operation of particle detectors in a HEP experiment results in exposure to radiation, which causes aging of detector components, referred to as radiation damage. Due to the difference in microscopic origin and macroscopic effect, one distinguishes between damage caused by ionizing and non-ionizing radiation, and the magnitude of exposure is expressed as a Total Ionizing Dose (TID, measured in \SI{}{\gray}) or Non-Ionizing Energy Loss (NIEL, expressed as a \SI{1}{\mega\electronvolt} neutron equivalent fluence in \SI{}{\neq\per\square\centi\meter}), respectively. The expected magnitude differs between proposals for future lepton colliders and depends on pending design choices, but is significantly smaller than expectations for the High Luminosity LHC~\cite{hllhc}. A vertex detector at FCC-ee operated at the Z pole, for example, would be required to tolerate a fluence of $\mathcal{O}(\SI{e13}{\neq\per\square\centi\meter})$ per year, and a TID of $\mathcal{O}(\SI{10}{\kilo\gray})$ per year~\cite{fccee}. For the other operation scenarios, the exposure is expected to be significantly lower, so that the total accumulated exposure, assuming a few years of operation at the Z pole, is expected to be a fluence of $\mathcal{O}(\SI{e14}{\neq\per\square\centi\meter})$ and a TID of $\mathcal{O}(\SI{100}{\kilo\gray})$.

The measurements on n-gap APTS in~\cite{apts} show noise below $30\,\text{e}^{-}$, $T_{99}$ above $150\,\text{e}^{-}$, and a deterioration of the spatial resolution of less than \SI{0.5}{\micro\meter} even after a TID of \SI{3}{\mega\gray} or a fluence of \SI{e14}{\neq\per\square\centi\meter} and higher fluences for pixel pitches of \SI{15}{\micro\meter} and \SI{10}{\micro\meter}. The measurements on DPTS~\cite{dpts} (also n-gap), indicate a degradation of the hit-detection efficiency at a fluence of \SI{e15}{\neq\per\square\centi\meter}, and an increased fake-hit rate (noise) after a TID of \SI{100}{\kilo\gray}. The performance at the next lower exposures, a fluence of \SI{e14}{\neq\per\square\centi\meter} or a TID of \SI{10}{\kilo\gray} is still acceptable, with a hit-detection efficiency of about \SI{99}{\percent} and negligible deterioration of the spatial resolution at a threshold of $125\,\text{e}^{-}$. This difference might be explained by a different level of circuit complexity, which render the DPTS more sensitive to TID exposure. The MOSS~\cite{moss} was characterized after exposure to a fluence of \SI{e13}{\neq\per\square\centi\meter} or a TID of \SI{10}{\kilo\gray}. The deterioration of the spatial resolution is negligible, and the deterioration of $T_{99}$ is below $25\,\text{e}^{-}$, while the fake-hit rate increases by up to two orders of magnitude.

\section{Simulations}
\label{simulations}
In the development of MAPS, simulations represent an essential complement to prototype characterization. Beyond serving as a key tool for understanding the behavior of existing sensors, they enable the prediction of the performance of different sensor layouts and geometries. In this context, simulations can significantly reduce both costs and development time by limiting the number of production cycles required to meet target specifications and by informing design decisions.

The presence of a strongly non-linear electric field significantly complicates the prediction of charge-collection dynamics in MAPS with a small collection electrode, which makes an accurate performance estimate difficult. In addition, the use of commercial processes for MAPS fabrication implies a lack of publicly available information on the manufacturing process and hence exact doping concentrations and profiles, which further complicates the prediction of sensor behavior.

Several studies have shown that it is possible to overcome these limitations by developing technology-independent simulation approaches~\cite{maps_sim_spannagel,maps_sim_wennolf}. These approaches, which are based on the fundamental principles of silicon detectors and utilize generic doping profiles, can describe and predict MAPS performance parameters through a combination of finite-element simulations and Monte Carlo methods. An overview of the simulation chain and relevant examples is given in the following. Further details on methods and procedures can be found in~\cite{maps_sim_wennolf}.

\subsection{Finite-element simulations}
Finite-element method simulations are a tool to model the electrostatic properties of and charge-transport in semiconductor devices. These properties are calculated at each node of a defined mesh by solving Poisson’s and carrier continuity equations.

Simulations can be performed in different modes. Quasi-stationary simulations are used to determine the sensor’s voltage-dependent electrical properties (electric field distribution, charge density, depletion regions, etc.). Transient simulations, on the other hand, are used to study the temporal response of the sensor to minimum ionizing particles.

In this context, Synopsys Sentaurus TCAD (Technology Computer-Aided Design)~\cite{synopsys_tcad_2025} and Silvaco TCAD~\cite{Silvaco} are commonly used tools for finite-elements simulations.

\subsection{Monte Carlo simulations}
Monte Carlo simulations are used to study the sensor response to incident particles, taking stochastic effects, like the variation of the energy deposition, into account. These cannot be accurately modeled directly with finite-element simulations due to the long computation time required for each event, which permits to obtain the required statistical significance.

In this context, the Allpix$^2$ framework~\cite{Allpix2} and Garfield++~\cite{garfieldpp} are among the most widely used tools. They allow the import of finite-element simulation results, such as electric fields and doping profiles, which are then mapped onto the sensor geometry. Charge creation can be simulated either directly or through interfaces to Geant4~\cite{Geant4_1,Geant4_2,geant4_3}. Charge carriers are subsequently propagated using non-transient or transient methods, with configurable mobility models, and the induced signals can be calculated.

\subsection{Front-end simulations}
Simulation of the sensor alone does not provide a complete picture of the detector's behavior, as the response is affected also by the readout electronics. TCAD and Monte Carlo tools are very accurate in describing charge generation and transport, but they do not reflect the characteristics of complex front-end stages, which often exhibit non-linear behavior due to their feedback and shaping mechanisms.

To simulate these effects, the current pulses produced by the sensor simulations can be used as an input to circuit simulations, to model the response and effects of the readout electronics. This approach allows the entire signal path to be reconstructed, including the amplitude, timing, and distortion introduced by the electronics. Tools commonly used for this purpose include Cadence Spectre~\cite{cadence_spectre} and LTspice~\cite{ltspice}.

\subsection{Simulation examples}
Several examples of simulations are available reproducing results from prototypes produced in the TPSCo \SI{65}{\nano\meter} ISC technology, which are discussed in this work. The simulated data are analyzed to extract key parameters such as hit-detection efficiency, cluster size, and spatial resolution. These can be directly compared with experimental measurements to validate the simulation model and investigate sensor characteristics.

In~\cite{apts_desy_sim}, test-beam measurements on APTS prototypes are compared with TCAD + Allpix$^2$ simulations, showing performance parameters, like cluster charge and hit detection efficiency, in close agreement with experimental data. In~\cite{maps_sim_viera}, transient simulations of APTS are used to compare the TCAD + Allpix$^2$ approach with TCAD-only simulations. It is shown that the transient simulation results from TCAD, in worst and best-case scenarios, can be reproduced by combining it with Allpix$^2$. This opens the possibility of faster transient studies for all possible injection positions, while also enabling the inclusion of stochastic effects, such as Landau fluctuations, thereby yielding more realistic and physically accurate simulation scenarios. In~\cite{maps_sim_mendes}, the studies are further extended by including SPICE (Simulation Program with Integrated Circuit Emphasis) tools in the simulation chain. Thanks to an accurate simulation of the front-end response, this approach allows the signal at the front-end output to be reproduced, yielding agreement between the peak positions of the $^{55}$Fe spectra from data and simulations. Further simulation validation with $^{55}$Fe spectra obtained using APTS prototypes is described in~\cite{maps_sim_sanna}, where Garfield++ is used. The results show good agreement with experimental data and Allpix$^2$ simulations, while discrepancies at low collected charge underline the importance of accurately modeling recombination and transport effects.

Another example for the importance of simulations is given by the simulation studies on H2M described in~\cite{h2m,h2m_sim}. It is shown that, the combination of TCAD, Monte Carlo, and front-end circuit simulations describes an unexpected asymmetric efficiency pattern observed in test-beam data~\cite{h2m,h2m_meas} (see also section~\ref{results:mip:eff}). The simulations suggest that the behavior originates from the weak lateral electric field, locally affected by the n-wells in the in-pixel circuitry. This slows charge collection which leads to a reduction of the signal amplitude at the output of the fast front end. This integrated simulation approach is able to qualitatively reproduce the measured in-pixel efficiency and provides guidance for mitigation strategies through layout adaptions in future designs.

The studies presented in~\cite{corentin} employ TCAD simulations to investigate variations of the low-dose n-implant, proposing a new version of the n-gap layout. This new version increases the doping concentration of the low-dose n-implant under the deep p-well. The simulations indicate that this improves the tolerance to NIEL.

In most of the presented studies, simulations were performed without proprietary information, suggesting that a technology-independent approach is often sufficient to describe primary sensor behavior. This is not the case for the studies in~\cite{maps_sim_sanna} and~\cite{corentin}, which rely on proprietary data, and the studies in~\cite{h2m,h2m_sim} further show that proprietary information may be essential to model specific sensor behavior.

\section{Conclusion and Outlook}
\label{conclusion}
This work summarizes the efforts to develop and characterize sensors for tracking and vertexing applications produced in the TPSCo \SI{65}{\nano\meter} ISC technology. The following section discusses the results presented above, from the point of view of an experimental application at a future lepton collider---as foreseen in the scope of the OCTOPUS project.
\paragraph{Spatial hit resolution} The OCTOPUS project aims for a spatial hit resolution better than \SI{3}{\micro\meter}. The presented results indicate, that this imposes a constraint on the pixel pitch, which depends on the employed sensor layout. For the n-gap layout this would have to be below \SI{15}{\micro\meter}, and below \SI{20}{\micro\meter} for the standard layout. Results on the n-blanket layout are relatively scarce, but the required pitch is expected to be between those of the other two sensor layouts. Furthermore, the results for a pitch of \SI{15}{\micro\meter} indicate that the spatial resolution attainable with the standard layout depends on the availability of charge information. If charge information is available, the spatial resolution can be $\mathcal{O}(\SI{1}{\micro\meter})$ better, depending on the reverse bias voltage.
\paragraph{Hit-detection efficiency and fake-hit rate} A hit-detection efficiency above \SI{99}{\percent} is achieved with a hit-detection threshold around $100\,\text{e}^{-}$ for all sensor layouts and most of the tested operation conditions. The operational margins for the n-blanket and n-gap layouts are still successively higher than for the standard layout. It is observed that noise and threshold dispersion, and consequently the fake-hit rate at a given threshold, vary significantly between prototypes, while hit-detection efficiency and spatial resolution profit from low thresholds. Thus, achieving a threshold level on the order of $100\,\text{e}^{-}$ should be a high priority.
\paragraph{Temporal hit resolution} The attainable temporal resolution is not as well studied as the previously discussed observables. The performance strongly depends on the sensor layout, pixel pitch, and is in some cases limited by the front end. The results from DPTS~\cite{dpts} indicate that the envisioned resolution on the order of \SI{5}{\nano\second} will most likely require time-walk corrections and has to be demonstrated for pixel pitches larger than \SI{15}{\micro\meter} and the selected sensor layout.
\paragraph{Material budget} Given the active thickness below \SI{10}{\micro\meter}, thinning to an overall thickness well below \SI{50}{\micro\meter} is possible, as demonstrated in~\cite{h2m}. This means that the sensors contribution to the material budget could be well below \SI{0.05}{\percent} of a radiation length $X_0$.
\paragraph{Power consumption} The OCTOPUS project aims for a total power consumption below \SI{50}{\milli\watt\per\square\centi\meter}. This depends heavily on the design of the analog front end and the readout architecture. At the time of writing, a front end similar to the DPTS design~\cite{DPTS_FE}, and an asynchronous readout architecture~\cite{async} are considered. The power consumption of the chosen front-end design is reported to be within the project requirement if the operation point is chosen appropriately.
\paragraph{Radiation hardness} Considering a few years of operation at the Z pole, a vertex detector at FCC-ee would be exposed to a fluence of $\mathcal{O}(\SI{e14}{\neq\per\square\centi\meter})$ and a TID of $\mathcal{O}(\SI{100}{\kilo\gray})$. The available measurements indicate that MAPS produced in the TPSCo \SI{65}{\nano\meter} ISC technology are likely to cope with this neutron equivalent fluence, if the n-gap layout is employed. Other sensor layouts are expected to be more sensitive to this type of irradiation. The expected TID will require care in the circuit design, to render it robust with respect to the change of transistor behavior. This may manifest in increased noise, as the presented studies suggest. In that regard, the reader is referred to~\cite{trans}, where measurements on irradiated transistor test structures produced in this technology are discussed.
\paragraph{Simulation} The presented simulation methods and studies are the core for the simulation efforts within the OCTOPUS project. A comprehensive and systematic study of sensor layout and geometry parameters is currently underway, combining finite-element, Monte Carlo, and front-end simulations. This will facilitate the efficient development of the optimal sensor and front end, aiming to address the ambitious project targets in terms of sensor performance. 
\\ \\ Overall, the specifications defined within the OCTOPUS project are in reach for a detector design employing the investigated process. Meeting all of them simultaneously will require making cautious design choices and finding a good compromise between competing properties. To facilitate this, the design process is accompanied by a thorough simulation campaign and supported by the data presented in this summary.

\section{Acknowledgments}
This work is heavily based on the design and characterization efforts of everyone working towards high-energy physics applications of the TPSCo \SI{65}{\nano\meter} ISC technology. The authors would like to express special gratitude towards Francesca Carnesecchi, Alessandra Lorenzetti, Eduardo Ploerer, Umberto Savino, Adriana Simancas, and Miljenko \v{S}ulji\'{c} for providing the data presented in the reproduced figures and overview plots and for providing additional information and clarification. Furthermore, we thank the ALICE ITS3 project in general for their openness to collaboration.

This work has been performed in the framework of the OCTOPUS project within the DRD3 Collaboration.

\section{CRediT author statement}
\textbf{Finn King:} Methodology, Investigation, Data Curation, Writing - Original Draft, Visualization 
\textbf{Matthew Lewis Franks:} Methodology, Investigation, Writing - Original Draft 
\textbf{Yajun He:} Methodology, Investigation, Writing - Original Draft 
\textbf{Gianpiero Vignola:} Investigation, Writing - Original Draft 
\textbf{Simon Spannagel:} Conceptualization, Writing - Review \& Editing, Project administration, Funding acquisition
\textbf{Malte Backhaus:} Writing - Review \& Editing 
\textbf{Auguste Besson:} Writing - Review \& Editing, Funding acquisition 
\textbf{Dominik Dannheim:} Conceptualization, Writing - Review \& Editing, Project administration 
\textbf{Andrei Dorokhov:} Writing - Review \& Editing 
\textbf{Ingrid-Maria Gregor:} Funding acquisition 
\textbf{Fadoua Guezzi-Messaoud:} Writing - Review \& Editing 
\textbf{Lennart Huth:} Investigation, Writing - Review \& Editing, Visualization 
\textbf{Armin Ilg:} Investigation, Writing - Review \& Editing 
\textbf{Zdenko Janoska:} Writing - Review \& Editing 
\textbf{Monika Kuncova:} Writing - Review \& Editing 
\textbf{Anna Macchiolo:} Project administration 
\textbf{Frédéric Morel:} Writing - Review \& Editing 
\textbf{Sara Ruiz Daza:} Investigation, Writing - Review \& Editing 
\textbf{Roberto Russo:} Investigation 
\textbf{Judith Schlaadt:} Writing - Review \& Editing 
\textbf{Serhiy Senyukov:} Writing - Review \& Editing 
\textbf{Peter Švihra:} Writing - Review \& Editing, Project administration, Funding acquisition 
\textbf{Anastasiia Velyka:} Investigation, Writing - Review \& Editing 
\textbf{Håkan Wennlöf:} Investigation, Writing - Review \& Editing

\appendix
\section{Tabulated data}
\label{app:tab}
The numbers presented in the Figures~\ref{fig:eff_mega}, \ref{fig:res_bin}, and~\ref{fig:res_ana}, were provided by the authors of the referenced publications. For better readability, the origin of the individual data points is not indicated in the figures, but summarized in tabular~\ref{tab:eff} and~\ref{tab:res}. If the provided data points did not contain the desired efficiencies or thresholds, the values were obtained from a linear interpolation between the closest data points. Additional caveats that need to be considered are discussed in the following.

The results for DPTS were obtained from a prototype irradiated to a dose of \SI{10}{\kilo\gray} and a \SI{1}{\mega e \volt} neutron equivalent fluence of \SI{e13}{\per\square\centi\meter}. The quoted $T_{99}$ is in agreement with the results from APTS at a reverse bias voltage of \SI{3.6}{\volt}, but significantly smaller than other measurement results for the same pitch. The spatial resolution is about \SI{0.8}{\micro\meter} larger than measurements on other prototypes of the same pitch suggest.

Some quoted spatial resolutions $\sigma_{\text{ana}}$ were corrected for the above-mentioned bias introduced through non-linear charge sharing ($\eta$-correction). Those are~\cite{apts} and \cite{apts-timing}. Furthermore, different authors use different methods to quantify the width of the spatial residual. The root mean square (RMS) or a truncated RMS are used in~\cite{apts,apts_desy_sim,dis_adri,h2m}, while the fit of a Gaussian function is used in~\cite{dpts}.

The spatial resolution and cluster size quoted for the H2M prototype were measured at a threshold of about $180\,\text{e}^{-}$, instead of the $100\,\text{e}^{-}$ at which other results were obtained.

\begin{table*}
    %\begin{scriptsize}
    \centering
    \begin{tabular}{ccccccc}
Prototype & Sensor layout   & Pitch [\SI{}{\micro\meter}] & Bias [V] & $T_{99}$ [$\text{e}^{-}$] & error +/- [$\text{e}^{-}$] & Reference\\
\midrule
APTS      & n-gap     &       10.0 &      1.2 &   159.4 &     1.3/ 7.5  & \cite{apts} \\
        \cmidrule(lr){3-7}
APTS      & n-gap     &       15.0 &      0.0 &   201.0 &     2.3/ 2.3  & \cite{apts} \\
APTS      & n-gap     &       15.0 &      1.2 &   184.1 &     2.5/ 6.3  & \cite{apts} \\
APTS      & n-gap     &       15.0 &      2.4 &   175.9 &     1.9/ 7.6  & \cite{apts} \\
APTS      & n-gap     &       15.0 &      3.6 &   155.9 &    18.6/ 6.8  & \cite{apts} \\
APTS      & n-gap     &       15.0 &      4.8 &   184.1 &     6.4/ 7.2  & \cite{apts} \\
        \cmidrule(lr){3-7}  
APTS      & n-gap     &       20.0 &      1.2 &   192.5 &     2.5/ 1.2  & \cite{apts} \\
        \cmidrule(lr){3-7}  
APTS      & n-gap     &       25.0 &      1.2 &   209.1 &     5.0/ 1.3  & \cite{apts} \\
APTS      & n-gap     &       25.0 &      1.2 &   210.4 &     0.1/ 0.1  & \cite{apts_desy_sim} \\
APTS      & n-gap     &       25.0 &      2.4 &   221.0 &     0.2/ 0.2  & \cite{apts_desy_sim} \\
APTS      & n-gap     &       25.0 &      3.6 &   203.5 &     0.1/ 0.1  & \cite{apts_desy_sim} \\
APTS      & n-gap     &       25.0 &      4.8 &   204.4 &     0.1/ 0.2  & \cite{apts_desy_sim} \\
    \cmidrule(lr){2-7}  
APTS      & n-blanket &       10.0 &      1.2 &   166.1 &     7.7/ 3.8  & \cite{apts} \\
        \cmidrule(lr){3-7}  
APTS      & n-blanket &       15.0 &      0.0 &   180.2 &     4.5/ 6.8  & \cite{apts} \\
APTS      & n-blanket &       15.0 &      1.2 &   179.0 &     3.8/ 1.3  & \cite{apts} \\
APTS      & n-blanket &       15.0 &      2.4 &   172.9 &     1.0/ 5.8  & \cite{apts} \\
        \cmidrule(lr){3-7}  
APTS      & n-blanket &       25.0 &      1.2 &   145.2 &     2.5/ 1.3  & \cite{apts} \\
APTS      & n-blanket &       25.0 &      1.2 &   149.8 &     0.1/ 0.1  & \cite{dis_adri} \\
    \cmidrule(lr){2-7}  
APTS      & standard  &       15.0 &      0.0 &    88.0 &     1.0/ 4.0  & \cite{apts} \\
APTS      & standard  &       15.0 &      1.2 &   103.0 &     0.9/ 1.8  & \cite{apts} \\
APTS      & standard  &       15.0 &      2.4 &   107.1 &     1.7/ 3.4  & \cite{apts} \\
APTS      & standard  &       15.0 &      3.6 &   106.9 &     4.1/ 3.3  & \cite{apts} \\
APTS      & standard  &       15.0 &      4.8 &   119.6 &     3.2/ 3.2  & \cite{apts} \\
        \cmidrule(lr){3-7}  
APTS      & standard  &       25.0 &      2.4 &   115.1 &     0.0/ 0.0  & \cite{dis_adri} \\
APTS      & standard  &       25.0 &      4.8 &   122.9 &     0.1/ 0.2  & \cite{dis_adri} \\
\midrule    
APTS-OA   & n-gap     &       10.0 &      1.2 &   153.0 &     2.0/ 2.0  & \cite{apts-timing} \\
APTS-OA   & n-gap     &       10.0 &      2.4 &   152.0 &     2.0/ 2.0  & \cite{apts-timing} \\
APTS-OA   & n-gap     &       10.0 &      3.6 &   147.0 &    11.0/11.0  & \cite{apts-timing} \\
APTS-OA   & n-gap     &       10.0 &      4.8 &   162.0 &     2.0/ 2.0  & \cite{apts-timing} \\
    \cmidrule(lr){2-7}  
APTS-OA   & n-blanket &       10.0 &      1.2 &   150.0 &    12.0/12.0  & \cite{apts-timing} \\
APTS-OA   & n-blanket &       10.0 &      2.4 &   155.0 &    12.0/12.0  & \cite{apts-timing} \\
APTS-OA   & n-blanket &       10.0 &      3.6 &   147.0 &    13.0/13.0  & \cite{apts-timing} \\
APTS-OA   & n-blanket &       10.0 &      4.8 &   158.0 &     5.0/ 5.0  & \cite{apts-timing} \\
\midrule
CE65v2    & n-gap     &       15.0 &      4.0 &   168.0 &     2.0/ 2.0  & \cite{ce65v2_pixel} \\
CE65v2    & n-gap     &       15.0 &     10.0 &   170.0 &     2.0/ 2.0  & \cite{ce65v2_pixel} \\
        \cmidrule(lr){3-7}  
CE65v2    & n-gap     &       22.5 &      4.0 &   165.0 &     2.0/ 2.0  & \cite{ce65v2_pixel} \\
CE65v2    & n-gap     &       22.5 &     10.0 &   156.0 &     2.0/ 2.0  & \cite{ce65v2_pixel} \\
    \cmidrule(lr){2-7}  
CE65v2    & standard  &       15.0 &      4.0 &   138.0 &     2.0/ 2.0  & \cite{ce65v2_pixel} \\
CE65v2    & standard  &       15.0 &     10.0 &   152.0 &     2.0/ 2.0  & \cite{ce65v2_pixel} \\
        \cmidrule(lr){3-7}  
CE65v2    & standard  &       22.5 &      4.0 &   120.0 &     2.0/ 2.0  & \cite{ce65v2_pixel} \\
CE65v2    & standard  &       22.5 &     10.0 &   136.0 &     2.0/ 2.0  & \cite{ce65v2_pixel} \\
\midrule
DPTS      & n-gap     &       15.0 &      1.8 &   156.2 &     2.7/ 2.7  & \cite{dpts} \\
DPTS      & n-gap     &       15.0 &      2.4 &   154.3 &     6.4/ 6.5  & \cite{dpts} \\
DPTS      & n-gap     &       15.0 &      3.0 &   159.1 &     6.9/ 7.0  & \cite{dpts} \\
\midrule
H2M       & n-gap     &       35.0 &      1.2 &   205.0 &     6.0/ 6.0  & \cite{h2m} \\
H2M       & n-gap     &       35.0 &      2.4 &   208.0 &     6.0/ 6.0  & \cite{h2m} \\
H2M       & n-gap     &       35.0 &      3.6 &   208.0 &     6.0/ 6.0  & \cite{h2m} \\
    \end{tabular}
    \caption{Tabulated version of the data presented in Figure~\ref{fig:eff_mega}. Prototype name and reference are supplementary. All results have been rounded to the same amount of digits, 0.0 means that the value is smaller than 0.05.}
    \label{tab:eff}
    %\end{scriptsize}
\end{table*}

\begin{table*}
    %\begin{scriptsize}
    \centering
    \begin{tabular}{cccccccc}
Prototype & Sensor layout & Pitch [\SI{}{\micro\meter}] & Bias [V] & $\sigma_{\text{bin}}$ [\SI{}{\micro\meter}] & $\sigma_{\text{ana}}$ [\SI{}{\micro\meter}] & cluster size & Reference \\
\hline
APTS      & n-gap     &       10.0 &      1.2 & 1.96 $\pm$ 0.01 & 1.92 $\pm$ 0.01 & 1.52             & \cite{apts} \\
        \cmidrule(lr){3-8}
APTS      & n-gap     &       15.0 &      1.2 & 3.11 $\pm$ 0.02 & 3.07 $\pm$ 0.02 & 1.47 $\pm$ 0.01  & \cite{apts} \\
APTS      & n-gap     &       15.0 &      2.4 & 3.18 $\pm$ 0.02 & 3.13 $\pm$ 0.02 & 1.45 $\pm$ 0.01  & \cite{apts} \\
APTS      & n-gap     &       15.0 &      3.6 & 3.10 $\pm$ 0.04 & 3.06 $\pm$ 0.04 & 1.46 $\pm$ 0.02  & \cite{apts} \\
APTS      & n-gap     &       15.0 &      4.8 & 3.16 $\pm$ 0.03 & 3.13 $\pm$ 0.03 & 1.42 $\pm$ 0.01  & \cite{apts} \\
        \cmidrule(lr){3-8}
APTS      & n-gap     &       20.0 &      1.2 & 4.41 $\pm$ 0.01 & 4.38 $\pm$ 0.01 & 1.39 $\pm$ 0.00 & \cite{apts} \\
        \cmidrule(lr){3-8}
APTS      & n-gap     &       25.0 &      1.2 & 5.82 $\pm$ 0.02 & 5.81 $\pm$ 0.02 & 1.34 $\pm$ 0.01 & \cite{apts} \\
APTS      & n-gap     &       25.0 &      1.2 &               - & 6.35 $\pm$ 0.05 & 1.47 $\pm$ 0.01 & \cite{apts_desy_sim} \\
APTS      & n-gap     &       25.0 &      2.4 &               - & 6.23 $\pm$ 0.11 & 1.48 $\pm$ 0.01 & \cite{apts_desy_sim} \\
APTS      & n-gap     &       25.0 &      3.6 &               - & 6.34 $\pm$ 0.05 & 1.47 $\pm$ 0.01 & \cite{apts_desy_sim} \\
APTS      & n-gap     &       25.0 &      4.8 &               - & 6.35 $\pm$ 0.08 & 1.45 $\pm$ 0.01 & \cite{apts_desy_sim} \\
    \cmidrule(lr){2-8}
APTS      & n-blanket &       10.0 &      1.2 & 1.88 $\pm$ 0.04 & 1.86 $\pm$ 0.04 & 1.54 $\pm$ 0.02 & \cite{apts} \\
        \cmidrule(lr){3-8}
APTS      & n-blanket &       15.0 &      1.2 & 2.79 $\pm$ 0.04 & 2.70 $\pm$ 0.04 & 1.65 $\pm$ 0.02 & \cite{apts} \\
APTS      & n-blanket &       15.0 &      2.4 & 2.90 $\pm$ 0.04 & 2.82 $\pm$ 0.04 & 1.58 $\pm$ 0.02 & \cite{apts} \\
        \cmidrule(lr){3-8}
APTS      & n-blanket &       25.0 &      1.2 & 4.30 $\pm$ 0.02 & 3.80 $\pm$ 0.02 & 2.00 $\pm$ 0.01 & \cite{apts} \\
APTS      & n-blanket &       25.0 &      1.2 &               - & 4.52 $\pm$ 0.04 & 2.25 $\pm$ 0.01 & \cite{dis_adri} \\
    \cmidrule(lr){2-8}
APTS      & standard  &       15.0 &      0.0 & 2.77 $\pm$ 0.03 & 2.34 $\pm$ 0.03 & 2.36 $\pm$ 0.02 & \cite{apts} \\
APTS      & standard  &       15.0 &      1.2 & 2.67 $\pm$ 0.03 & 2.30 $\pm$ 0.03 & 2.27 $\pm$ 0.02 & \cite{apts} \\
APTS      & standard  &       15.0 &      2.4 & 2.51 $\pm$ 0.03 & 2.08 $\pm$ 0.02 & 2.19 $\pm$ 0.02 & \cite{apts} \\
APTS      & standard  &       15.0 &      3.6 & 2.58 $\pm$ 0.03 & 2.16 $\pm$ 0.02 & 2.18 $\pm$ 0.02 & \cite{apts} \\
APTS      & standard  &       15.0 &      4.8 & 2.48 $\pm$ 0.03 & 2.08 $\pm$ 0.02 & 2.11 $\pm$ 0.02 & \cite{apts} \\
        \cmidrule(lr){3-8}
APTS      & standard  &       25.0 &      2.4 &               - & 4.99 $\pm$ 0.01 & 2.81 $\pm$ 0.01 & \cite{dis_adri} \\
APTS      & standard  &       25.0 &      4.8 &               - & 4.14 $\pm$ 0.06 & 2.70 $\pm$ 0.01 & \cite{dis_adri} \\
\midrule
APTS-OA   & n-gap     &       10.0 &      1.2 &               - & 1.91 $\pm$ 0.01 & 1.53 $\pm$ 0.75 & \cite{apts-timing} \\
APTS-OA   & n-gap     &       10.0 &      2.4 &               - & 2.01 $\pm$ 0.01 & 1.47 $\pm$ 0.71 & \cite{apts-timing} \\
APTS-OA   & n-gap     &       10.0 &      3.6 &               - & 2.01 $\pm$ 0.02 & 1.43 $\pm$ 0.68 & \cite{apts-timing} \\
APTS-OA   & n-gap     &       10.0 &      4.8 &               - & 2.03 $\pm$ 0.01 & 1.45 $\pm$ 0.69 & \cite{apts-timing} \\
    \cmidrule(lr){2-8}
APTS-OA   & n-blanket &       10.0 &      1.2 &               - & 1.85 $\pm$ 0.02 & 1.53 $\pm$ 0.74 & \cite{apts-timing} \\
APTS-OA   & n-blanket &       10.0 &      2.4 &               - & 1.98 $\pm$ 0.02 & 1.49 $\pm$ 0.72 & \cite{apts-timing} \\
APTS-OA   & n-blanket &       10.0 &      3.6 &               - & 2.05 $\pm$ 0.02 & 1.47 $\pm$ 0.69 & \cite{apts-timing} \\
APTS-OA   & n-blanket &       10.0 &      4.8 &               - & 2.06 $\pm$ 0.01 & 1.45 $\pm$ 0.69 & \cite{apts-timing} \\
\midrule
CE65v2    & n-gap     &       15.0 &      4.0 &               - & 3.13 $\pm$ 0.02 &               - & \cite{ce65v2_pixel} \\
CE65v2    & n-gap     &       15.0 &     10.0 &               - & 3.14 $\pm$ 0.02 &               - & \cite{ce65v2_pixel} \\
        \cmidrule(lr){3-8}
CE65v2    & n-gap     &       22.5 &      4.0 &               - & 5.12 $\pm$ 0.02 &               - & \cite{ce65v2_pixel} \\
CE65v2    & n-gap     &       22.5 &     10.0 &               - & 4.84 $\pm$ 0.01 &               - & \cite{ce65v2_pixel} \\
    \cmidrule(lr){2-8}
CE65v2    & standard  &       15.0 &      4.0 &               - & 1.40 $\pm$ 0.02 &               - & \cite{ce65v2_pixel} \\
CE65v2    & standard  &       15.0 &     10.0 &               - & 1.91 $\pm$ 0.02 &               - & \cite{ce65v2_pixel} \\
        \cmidrule(lr){3-8}
CE65v2    & standard  &       22.5 &      4.0 &               - & 1.94 $\pm$ 0.02 &               - & \cite{ce65v2_pixel} \\
CE65v2    & standard  &       22.5 &     10.0 &               - & 2.11 $\pm$ 0.02 &               - & \cite{ce65v2_pixel} \\
\midrule
DPTS      & n-gap     &       15.0 &      2.4 & 3.88 $\pm$ 0.03 &               - & 1.37 $\pm$ 0.01 & \cite{dpts} \\
DPTS      & n-gap     &       15.0 &      3.0 & 3.91 $\pm$ 0.03 &               - & 1.38 $\pm$ 0.01 & \cite{dpts} \\
\midrule
H2M       & n-gap     &       35.0 &      1.2 & 9.30 $\pm$ 0.10 &               - & 1.19 $\pm$ 0.01 & \cite{h2m} \\
    \end{tabular}
    \caption{Tabulated version of the data presented in Figure~\ref{fig:res_bin} and~\ref{fig:res_ana}. Prototype name and reference are supplementary. All results have been rounded to the same amount of digits, 0.00 means that the value is smaller than 0.005.}
    \label{tab:res}
    %\end{scriptsize}
\end{table*}

\clearpage
\bibliographystyle{elsarticle-num}
\bibliography{bibliography}

\end{document}